\PassOptionsToPackage{disable}{endfloat}
\documentclass[useAMS, referee, usenatbib]{biom}

\def\bSig\mathbf{\Sigma}

\usepackage[figuresright]{rotating}
\usepackage{mathtools}
\usepackage{comment}
\usepackage{float}
\usepackage{algorithm}
\usepackage{algpseudocode}
\usepackage{amsmath,amssymb}  
\usepackage{graphicx,psfrag,epsf}
\usepackage{booktabs} 
\usepackage{multirow} 
\usepackage{xcolor}
\usepackage{url}      
\usepackage{macros}

\usepackage{mathtools}  
\usepackage{amssymb}
\usepackage{amsfonts}
\usepackage{bm}
\usepackage{dsfont}

\usepackage{graphicx}
\usepackage{booktabs}
\usepackage{multirow}
\usepackage{threeparttable} 
\usepackage{subcaption}     
\usepackage{enumitem}  
\usepackage{multicol}   
\usepackage{adjustbox}

\usepackage{algorithm}
\usepackage{algpseudocode}

\usepackage{macros}

\usepackage{url}
\usepackage{natbib}
\usepackage[hidelinks,hypertexnames=false]{hyperref}
\usepackage{etoolbox}
\patchcmd{\endtabular}{\vspace*{-20pt}}{}{}{}
\expandafter\let\csname endtabular*\endcsname\endtabular
\makeatletter
\let\ps@titlepage\ps@empty
\makeatother
\def\cT{\mathcal{T}}
\def\cM{\mathcal{M}}

\title[Graph-Split Bayesian Causal Forest]{Graph-Split Bayesian Causal Forest for Spatial Heterogeneous Treatment Effect Estimation}

\author{Shuren He$^{1*}$,
Huiyan Sang$^{1*}$\email{huiyan@stat.tamu.edu}, and 
Ligang Lu$^{2}$ \\
$^{1}$Department of Statistics, Texas A\&M University, College Station, TX, USA \\
$^{2}$ Shell USA, Inc., Houston, TX, USA} 

\begin{document}









\label{firstpage}


\begin{abstract}
In spatial observational studies, treatment assignment and outcomes often exhibit spatial dependence patterns, and treatment effects may vary across space and subpopulations due to both measured and unmeasured spatially structured confounders. Accounting for spatial dependence while estimating heterogeneous treatment effects (HTEs) is a central task in spatial causal inference. Causal Bayesian additive regression tree methods are popular nonparametric methods for modeling and estimating HTEs. Despite their flexibility and uncertainty quantification, the axis-aligned split rules often adopted in these models are not suitable for modeling spatial structures.  
We propose a spatial structure–aware Bayesian nonparametric method, called Graph-Split Bayesian Causal Forest (GSBCF), that integrates graph-split Bayesian additive regression trees (GS-BART) with the Bayesian causal forest propensity-score regression framework for spatial heterogeneous causal inference. Spatial confounding is accommodated through graph-guided split rules in modeling decision trees of the prognostic and HTE functions. We develop an efficient informed proposal sampling algorithm for posterior computation, enabling full Bayesian inference of the spatial conditional average treatment effect function. 
Simulations and a real data study demonstrate substantially improved estimation accuracy and uncertainty quantification over existing causal BART methods.
\end{abstract}

%



\maketitle


%

\section{Introduction}
\label{sec:intro}
Spatial causal inference has attracted increasing attention in recent years and has emerged as a growing area of methodological research, with wide applications in biomedical and public health studies. Many spatial causal inference problems arise from observational studies, where geographical characteristics strongly influence both treatment assignment and outcomes. In such settings, explicitly incorporating spatial information is important to account for confounding effects due to both measured and unmeasured spatially structured factors~\citep{reich2021review}. Moreover, accounting for heterogeneous causal effects~\citep{athey2015machine,hahn2020bayesian} is particularly important in spatial contexts to quantify variation in treatment effects across locations and subpopulations.
  
Despite recent advances in spatial causal inference aimed at accounting for spatial dependence~\citep{gao2022causal,reich2021review}, the existing literature remains relatively limited, particularly with respect to modeling spatial HTEs. Most existing methods rely on conventional spatial (generalized) linear or additive models based on Gaussian processes (GP) or Markov random fields (MRF) for modeling outcomes,  propensity scores~\citep{papadogeorgou2019adjusting}, or HTEs~\citep{osama2019inferring}, and typically assume a low-dimensional set of preselected covariates and a constant treatment effect~\citep{reich2021review}. These limitations restrict their ability to capture complex confounding and HTE effects. 

Causal inference based on Bayesian additive regression trees (BART) models~\citep{hahn2020bayesian,krantsevich2023stochastic} has emerged as a popular Bayesian machine learning approach for heterogeneous causal effect estimation under the potential outcome framework, owing to its ability to flexibly capture complex nonlinear relationships and interactions, while providing posterior uncertainty quantification for causal estimands. However, existing causal BART and causal forest models mostly rely on axis-aligned tree partitions, making them inefficient to model latent functions with spatial structures and spatial domains with irregular geometry commonly seen in applications.  

\noindent\textbf{Contributions:} To address these limitations, we propose Graph-Split Bayesian Causal Forest (GSBCF), a spatial structure–aware Bayesian nonparametric framework for estimating spatially heterogeneous treatment effects. We
define spatial causal estimands and identify the required causal assumptions under the potential outcome framework. 
GSBCF integrates the recently proposed spatial nonparametric function prior based on graph-split Bayesian additive regression trees, called GS-BART \citep{he2026gs}, into the Bayesian causal forest propensity score regression framework for spatial heterogeneous treatment effect estimation. Specifically, GS-BART provides graph-guided tree partitioning that accounts for the joint effects of measured covariates and unmeasured spatial factors when modeling the spatial propensity score, prognostic function, and heterogeneous treatment effect function. 
The method replaces axis-aligned decision rule partitions with graph-guided decision rules, allowing for flexible spatial partitions while respecting spatial contiguity and domain topology. This enables flexible modeling of spatial confounding and heterogeneous causal effects without imposing restrictive smoothness or parametric assumptions. In contrast to GP or MRF-based approaches, GSBCF naturally accommodates spatial patterns with varying smoothness and nonlinear interactions between spatial structure and covariates.
To our knowledge, GSBCF is the first Bayesian causal forest to directly incorporate spatial structure into tree partitioning for heterogeneous treatment effect estimation. We develop an efficient informed sampling algorithm for varying-coefficient graph-split trees, enabling scalable Bayesian inference with calibrated uncertainty quantification. Through extensive simulations and a county-level COPD study, we demonstrate that GSBCF substantially improves estimation accuracy and uncertainty calibration over existing causal BART methods. Software implementing GSBCF is publicly available at \url{https://github.com/shuren-he/GSBCF}.


\section{Causal inference framework for spatial data}
\label{sec:causal-framework}
Let $Y_i$ denote the spatial response variable at spatial unit $i$, and  $Z_i \in \{0,1\}$ is a binary treatment variable. $\bfX_i\in \mathbb{R}^{d}$ represents a vector of the measured control variables for the $i$th observation. Here, a spatial unit may either represent an areal unit (e.g., a county) or a spatial point (e.g., a weather station). With a slight abuse of notation, we use $\bfS_i$ to represent the location information of the spatial unit $i$. It is reasonable to assume that there is a spatial effect on the outcome unexplained by observed covariates and possibly due to some missing spatial factors, denoted as $\bfU_i\coloneq \bfU(\bfS_i)$. We also assume that there is a spatial effect on treatment due to another set of missing spatial factors related to treatment, denoted as $\bfV_i\coloneq \bfV(\bfS_i)$. Note that $\bfU_i$ and $\bfV_i$ may have some shared spatial factors that can confound the treatment effects. 
We use lowercase Roman letters to denote the values assumed by variables. 
Our data consist of $n$ independent observations $(y_i, z_i, \bfx_i, \bfs_i)$. 

\noindent\underline{Spatial causal estimands.} We begin by defining spatial causal estimands under the potential outcome framework.  
For each unit $i=1,\ldots,n$,  let $Y_i(1)$ and $Y_i(0)$ be the potential outcomes under treatment and control, respectively. The observed outcome can be written as
$Y_i = Z_i Y_i(1) + (1-Z_i)Y_i(0)$.
We define the conditional average treatment effect (CATE) as
\begin{eqnarray}\label{eq:CATE}
\tau(\bfx,\bfs) := \mathbb{E}\!\left[Y(1)-Y(0)\mid \bfX=\bfx,\bfS=\bfs\right].
\end{eqnarray}
The spatially marginalized CATE is obtained by averaging the CATE over the covariate distribution: 
$\tau(\bfs) := \mathbb{E}_{\bfX}\big[\tau(\bfx,\bfs)\big]$.
Marginalized CATEs for other covariates can be defined in a similar fashion. Finally, we define the spatial sample averaged treatment effect (ATE) as: 
$\tau := \frac{1}{n}\sum_{i=1}^{n} \tau(\bfx_i,\bfs_i)$.

We adopt the following assumptions required for our spatial causal inference: 
\begin{itemize}
    \item \textbf{SUTVA (Stable Unit Treatment Value Assumption):} No treatment assignment to a particular individual should affect the observed outcomes on other individuals, and there is no variation in treatment. Under SUTVA, we assume no interference across spatial units. However, we allow for spatial dependence in both treatment assignment and outcomes through measured and unmeasured spatial confounders.
    \item \textbf{Spatial latent ignorability assumption:} $Y_i(z)$ and $Z_i$ are conditionally independent, i.e., 
    $  Y_i(0), Y_i(1) \perp Z_i \mid \bfX_i, \bfU_i, \bfV_i$.
    \item \textbf{Positivity:} 
    $ 0<\operatorname{Pr}\left(Z_i=1 \mid \bfX_i, \bfV_i \right)<1$. 
\end{itemize}
\noindent\underline{Remark on assumptions:}  In this paper, we focus on spatial application problems where the SUTVA assumption is reasonable. There are numerous such examples in practice. In agricultural field trials, plots are often separated by buffer zones to prevent fertilizer or irrigation spillover. In education policy studies, treatments are typically assigned at the school or district level (e.g., funding or curriculum changes) and primarily affect students within that unit. In shale oil production, engineering completion parameters are controlled at the individual well level, with minimal cross-well interference in production outcomes. In these settings, methods that assume SUTVA are both practically relevant and methodologically appropriate. For applications where spatial interference may occur, the current GSBCF framework can be extended, and we provide some possible ways in the Conclusion section. In addition, we evaluate the robustness of GSBCF through a simulation sensitivity study under mild violations of SUTVA in Section~\ref{subsec:sensi}.

The spatial latent ignorability is a standard assumption commonly adopted in the spatial causal inference literature \citep[see, e.g.,][]{reich2021review}. In spatial problems, important unmeasured factors, such as environmental exposures and socioeconomic conditions, often vary spatially and exhibit some degree of spatial dependence patterns. We introduce $\bfU(\bfS_i)$ and $\bfV(\bfS_i)$ to conceptually explain the different sources of spatial variation and confounding affecting the outcome and treatment assignment that are unexplained by the observed covariates. Their joint effects with the observed covariates on treatment assignment, the prognostic function, and treatment effects are the primary quantities of interest in model inference. Therefore, we shall model the effects of these measured and unmeasured spatial factors through flexible unknown functions of $(\bfX,\bfS)$, avoiding explicit modeling of the individual latent factors $\bfU$ and $\bfV$, their functional effects, and their potentially high-dimensional dependence structures. If latent confounders exhibit weak or no spatial structure, or if the spatial resolution is too coarse to capture small-scale spatial dependence, residual confounding may remain that would violate the spatial ignorability assumption~\citep{papadogeorgou2019adjusting,osama2019inferring,
reich2021review,gao2022causal}. In such settings, additional information, such as additional covariates, higher-resolution spatial data, or spatiotemporal replication, would be helpful. 

Finally, the positivity assumption requires sufficient overlap in the covariate distributions between treated and control units across space.
Under the model assumption that the effects of $\bfV$ can be fully
represented through functions of $\bfS$, conditioning
on $(\bfX,\bfS)$ would capture the relevant confounding information contained in
$(\bfX,\bfV)$. In particular, the positivity condition  has a corresponding observed-data form $
0<\Pr(Z_i=1\mid \bfX_i,\bfS_i)<1$.
Careful diagnostic checks of estimated propensity scores are essential to ensure the positivity assumption is approximately satisfied.

\noindent\underline{Spatial propensity score.} We introduce the notion of spatial propensity score, extending the classical propensity score to the spatial setting.  The propensity score at the $i$th spatial unit is defined as
$e_i=\Pr\{Z_i=1\mid \bfx_i,\bfs_i\}$,
which generalizes the usual propensity score by explicitly recognizing that treatment assignment may depend on both $\bfX_i$ and $\bfV_i$.  
A common treatment assignment model uses a logistic link: $Z_{i} \sim\text{Ber}(e_{i}) \text{ with }\text{logit}(e_{i})=\mu_{z}(\bfx_i,\bfs_i)$, where $\mu_{z}(\bfx_i,\bfs_i)$ is the logit-scale treatment assignment regression function, which captures the joint effects of the observed covariates $\bfx_i$ and
unobserved spatial confounders $\bfV_i$ on treatment assignment, without requiring
explicit modeling of each unobserved spatial variable in $\bfV_i$.
In standard spatial causal inference models, $\mu_{z}(\bfx_i,\bfs_i)$ is often assumed to take an additive form separating the effects of covariates and unmeasured spatial factors $\bfV_i$, $\mu_{z}(\bfx_i,\bfs_i)=\mu_{z}(\bfx_i)+\mu_{z}(\bfs_i)$, where $\mu_{z}(\bfx_i)$ is often modeled by a parametric regression model such as linear regression, and $\mu_{z}(\bfs_i)$ is often modeled by a spatial random effect through Gaussian process (GP) or Gaussian Markov random fields (GMRF). We shall relax this separability assumption in GSBCF by a more flexible spatially informed additive tree model in Section~\ref{sec:GS-BCF}. GSBCF also offers greater flexibility than GP- or GMRF-based spatial causal inference approaches, which often rely on common spatial dependence parameters, by capturing more complex spatial patterns such as sharp changes, spatially varying smoothness, and interactions with covariates. This flexibility may help mitigate residual confounding and better support the spatial latent ignorability assumption.


\noindent\underline{Spatial propensity score outcome regression model.} The estimated spatial propensity scores from the treatment assignment model, denoted as $\hat{e}_i$, can be used in various ways in observational studies, including matching~\citep{kane2020propensity}, weighting~\citep{li2018balancing}, and stratification~\citep{lunceford2004stratification}. We focus on the propensity score regression approach, as it allows the propensity score and additional covariates to be incorporated simultaneously into a unified outcome model to mitigate confounding bias and improve covariate balance between treated and control units~\citep{fan2021bayesian}.

Specifically, we consider a propensity score regression model with mean-zero normal errors for the outcome:
$
y_i=\mathrm{E}\left(Y_i \mid \bfX_i, \bfU_i, \hat{e}_i, Z_i\right)+\epsilon_i, \quad \epsilon_i \sim N\left(0, \sigma^2\right),
$
where we model 
\vspace{-2mm}
\begin{equation}\label{eq:outcomeE}
\mathrm{E}\left(Y_i \mid \bfX_i, \bfU_i,\hat{e}_i, Z_i\right)=\mu\left(\bfx_i, \bfs_i, \hat{e}_i\right)+\tau\left(\bfx_i, \bfs_i\right) (z_i-0.5).
\end{equation}
Here, $\mu(\cdot)$ is called the prognostic term capturing the baseline response surface, and $\tau(\cdot)$ represents the conditional treatment effect capturing the heterogeneous causal effects of $Z$. 
Indeed, under our causal assumptions, CATE is estimated from
$
\tau(\mathbf{x}_i,\bfs_i)=\mathbf{E}(Y \mid \bfx_i, \bfs_i,  Z_i=1)-\mathbf{E}(Y \mid \bfx_i,\bfs_i, Z_i=0) .
$
Combine with the formulation in~\eqref{eq:outcomeE}, the regression coefficient of $z_i$ in~\eqref{eq:outcomeE} directly corresponds to the spatial CATE defined in~\eqref{eq:CATE}.

We remark that both $\mu(\cdot)$ and $\tau(\cdot)$ are modeled as joint functions of
$\bfs$ and $\bfx$. Rather than explicitly modeling the latent spatial
confounders $\bfU$, their effects are absorbed through the spatial component
$\bfs$, avoiding high-dimensional latent variables while accounting for their
spatial structures and potential interactions with $\bfX$.

We include the estimated propensity score $\hat e$ only in the prognostic
function $\mu(\cdot)$ to adjust for confounding, and exclude it from the
treatment effect function $\tau(\cdot)$, as allowing treatment effects to depend on the treatment assignment mechanism may confound causal interpretation~\citep{hahn2020bayesian}.


Note that we also adopt a centering strategy to convert $z_i$ to $z_i-0.5$ for the treatment variable in~\eqref{eq:outcomeE}. Without centering, the prior variance of the control
outcome is governed by the prior on $\mu(\cdot)$, whereas the prior variance of
the treatment outcome equals the sum of the prior variances of $\mu(\cdot)$ and
$\tau(\cdot)$, inducing asymmetric prior dispersion. The centering strategy enables balanced prior variability across groups.

\section{Nonparametric prior on latent spatial regression functions by GS-BART}
Our spatial causal inference framework involves modeling the treatment
assignment function $\mu_z(\cdot)$, the prognostic function $\mu(\cdot)$, and
the treatment effect function $\tau(\cdot)$, all of which depend on spatial
location $\bfs$ and other predictors. To model a generic latent spatial
regression function $\phi(\bfx,\bfs)$, much of the spatial literature assumes
an additive decomposition $\phi(\bfx,\bfs)=\phi(\bfx)+\phi(\bfs)$, where
$\phi(\bfs)$ captures residual spatial dependence unexplained by the
covariates. GP, MRF, and spline-based methods are commonly used to model $\phi(\bfs)$. 
Various recent work combines machine learning models for $\phi(\bfx)$ with spatial random effect models for $\phi(\bfs)$ in separable
regression frameworks, including BART with CAR models~\citep{zhang2007spatially},
random forests with nearest-neighbor GPs~\citep{saha2023random}, LightGBM with
Vecchia GPs~\citep{sigrist2022gaussian}, and neural networks with GPs
~\citep{zhan2024neural}. However, these approaches inherit the limitations of
GP or MRF-based spatial random effect models that typically impose stringent stationarity and global smoothness assumptions. In addition, the
assumed separability between $\bfx$ and $\bfs$ may fail to capture their
interactions.

Very recently, \citet{he2026gs} proposed a GS-BART (Graph-Structured Bayesian Additive Regression Trees) model, which extends the standard BART model to handle features with graph structures in regression problems. Unlike the original BART, which relies on parallel and axis-aligned splitting constraints, GS-BART allows more flexible partitioning guided by the underlying graph topology, making it particularly well suited for modeling spatial data. 
We briefly review the GS-BART model below.

Let $[n]$ denote the observed index set. 
In GS-BART, we model $\phi(\bfx,\bfs)$ by an additive form of $T$ functions,
\begin{equation}\label{eq:addtivemodel}
    \phi(\bfx,\bfs)=\sum_{t=1}^T g_t(\bfx,\bfs;\mathcal{T}_t, \mathcal{M}_t) 
\end{equation}
where $g_t(\bfx,\bfs)$ is called the $t$th weak learner function parameterized by a decision tree $\mathcal{T}_t$ and leaf weight parameters $\mathcal{M}_t$. 
Each decision tree $\mathcal{T}$ recursively partitions the data into leaf nodes 
$\underline{\xi}(\mathcal{T}) = \{\xi_1(\mathcal{T}), \dots, \xi_\ell(\mathcal{T})\}$ according to decision rules at each internal node, forming a partition of $[n]$. 
Samples within a leaf node $\xi_k$ share a constant mean (leaf weight) $\mu_k \in \mathbb{R}$, and we collect all leaf parameters as $\mathcal{M} = (\mu_1, \dots, \mu_\ell)$. 


Conventional decision tree models often assume the inputs are a set of unstructured features, and the
decision rule at each internal tree node takes the form $x_j>c$ or $x_j<c$,
using only one covariate $x_j$ at a time. In GS-BART, each tree $\mathcal{T}_t$
is instead equipped with a set of $p_t$ candidate directed rooted graphs (arborescences) $\dbmG_t=\{\dG_{j}^{(t)}\}_{j=1}^{p_t}$ that encode graph structural information to
guide the partitioning process.
A graph $\dG$ has vertex set $V = \{v_1, \dots, v_{|V|}\}$ and directed edge set $\dE$, where $V$ induces a pre-specified partition of $[n]$ to bin the data for memory and training computational efficiency, and $\dE$ encodes feature similarity among vertices. 

Arborescences are adopted since removing any edge $\de\in\dE$ yields a
bipartition of $V$ into two connected components $V_L$ and $V_R$, and both
induced subgraphs remain arborescences. Consequently, this induces a bipartition of samples attached to each vertex. Therefore, this graph bipartition procedure can be considered as the split rule at an internal node, enabling recursive graph-based partitions. 

To model $\phi(\bfx,\bfs)$, we encode both covariate and spatial information
through the construction of $\dbmG_t$. For numerical
covariates, we sort the observations and bin adjacent values, following the
strategy used in LightGBM, and construct directed chain graphs (special cases of
arborescences) connecting the bins in order. Removing a decision edge in such a
graph is equivalent to a classical threshold split $X > c$, as in standard
BART. For spatial areal data, we construct candidate graphs by sampling random
arborescences from a spatial adjacency graph $A_0$. To reduce computational complexity, nodes are first binned, for example via community detection, and random spanning trees of $A_0$ are then oriented into arborescences. For point-referenced spatial data on a manifold $\calS$, we discretize $\calS$ into spatial bins (e.g., Voronoi tessellations), form a neighborhood graph $G_0$ by linking adjacent bins, and sample random arborescences of $G_0$ to include in
$\dbmG_t$. Each spatial graph-based split induces a contiguous bipartition of the spatial domain respecting domain geometry. When modeling only $\phi(\bfs)$, GS-BART includes only spatial
arborescences and models the spatial effect as a sum of $T$ spatial piecewise-constant functions.

Given $\dbmG_t$, we define a graph-split decision tree generative prior. Starting with the
root node representing $[n]$, each leaf node $\xi$ splits with probability
$p_{\text{split}}(\xi)=\alpha(1+d_\xi)^{-\beta}$, where $d_\xi$ is the depth of
$\xi$. 
Conditional on splitting, a \emph{decision arborescence} $\dG\in\dbmG_t$
and a \emph{decision edge} $\de\in\dE$ are selected from the set of valid rules
that yield nonempty children.   
For leaf parameters $\calM=(\mu_1(\calT),\ldots,\mu_\ell(\calT))$, we assume
$\mu_k\stackrel{\text{ind}}{\sim}N(\mu_0,\sigma_f^2)$. 

The function $\phi(\bfx,\bfs)$ modeled by GS-BART is a nonseparable function of $\bfx$ and
$\bfs$ whenever at least one tree includes both spatial arborescences and
covariate chain arborescences in its decision rules. This enables the model to
capture interactions between measured $\bfX$ and unmeasured spatial effects. GS-BART further allows the candidate graph set $\dbmG_t$ to vary across trees (e.g., through
different spatial tessellations for binning). Consequently, the fitted ensemble is not tied to a single spatial binning, and observations grouped within the same bin in one tree may be separated by alternative graph partitions or covariate splits in other trees. This enhances ensemble diversity and yields more expressive decision boundaries.
Due to the additive structure of the model, user-specified combinations of candidate graph sets can also be adopted to reflect prior assumptions on specific covariate or interaction effects.


\section{Spatial Causal Inference Method}
\label{sec:GS-BCF}
\subsection{Bayesian Spatial Causal Forest Model}
We now describe the Graph-Split Bayesian Causal Forest (GSBCF) model based on GS-BART model. 
The functions $\mu_z(\cdot)$, $\mu(\cdot)$ and $\tau(\cdot)$ are each represented as ensembles of regression trees:
\begin{align}
\mu_z(\mathbf{x}, \mathbf{s})
&= \sum_{t=1}^{T^{(e)}} g_{t}^{(e)}(\mathbf{x}, \mathbf{s};\mathcal{T}_t^{(e)}, \mathcal{M}_t^{(e)}),
\label{eq:ps-bart}
\\
\mu(\mathbf{x}, \hat{e}, \mathbf{s})
&= \sum_{t=1}^{T^{(\mu)}} g_{t}^{(\mu)}(\mathbf{x}, \hat{e}, \mathbf{s};\mathcal{T}_t^{(\mu)}, \mathcal{M}_t^{(\mu)}),
\label{eq:mu-bart}
\\
\tau(\mathbf{x}, \mathbf{s})
&= \sum_{t=1}^{T^{(\tau)}} g_{t}^{(\tau)}(\mathbf{x}, \mathbf{s};\mathcal{T}_t^{(\tau)}, \mathcal{M}_t^{(\tau)}),
\label{eq:tau-bart}
\end{align}
where $T^{(\cdot)}$ denotes the number of trees used in GS-BART to model each unknown function. 
Each $\mathcal{T}_t^{(\cdot)}$ represents an individual decision tree, and
$\mathcal{M}_t^{(\cdot)}$ is the corresponding vector of scalar means associated with
the leaf nodes of $\mathcal{T}_t^{(\cdot)}$. 
Each tree is associated with its own candidate graph set.  
We include all available measured covariates $\bfX$ in~\eqref{eq:ps-bart}, ~\eqref{eq:mu-bart}, and ~\eqref{eq:tau-bart} by default. However, different covariate subsets may be used to construct GS-BART candidate graphs when prior knowledge is available.  

We estimate the propensity scores using the generalized GS-BART
logistic regression model of \citet{he2026gs}, with treatment
assignment as the response and observed covariates and spatial
location as predictors. We use the posterior mean treatment
probabilities as $\hat e_i$ and include them only in the prognostic
forest, through an additional covariate chain graph.
Further estimation details are provided in the Supplementary
Section~\ref{appsub:propensity}.

 
Motivated by the belief that treatment effect heterogeneity is simpler than the
prognostic effect, we impose different depth penalties in the leaf-node split
prior for the prognostic and treatment trees, following \citet{hahn2020bayesian}. Specifically, we set $\alpha=0.95$ and $\beta=2$ for the prognostic trees. We enforce stronger regularization for the treatment trees by setting $\alpha=0.25$ and $\beta=3$.
The residual variance $\sigma^2$ is assigned an inverse-gamma prior. 
We assign a normal prior to the leaf weight parameters,
$\mathcal{M}_t^{(\mu)} \sim \mathcal{N}\left(\mathbf{0}, \sigma_{\mu}^2 I\right)$, and 
$\mathcal{M}_t^{(\tau)} \sim \mathcal{N}\left(\mathbf{0}, \sigma_{\tau}^2 I\right)$. 
In standard BART regression and causal inference models, the leaf weight prior variance is often fixed at a data-driven empirical value as recommended in ~\citet{chipman2010bart}. However, we note that the leaf-weight prior variance parameters, $\sigma_\mu^2$ and $\sigma_\tau^2$, also correspond to the prior variances of $\mu(\cdot)$ and $\tau(\cdot)$, respectively, after integrating out the leaf-weight parameters. They play an important role in controlling the overall function magnitude and hence regulating prior shrinkage and model regularization. Therefore, we place independent inverse-Gamma priors,
$\sigma_\mu^{2},\, \sigma_\tau^{2} \sim \text{Inv-Gamma}\left(\frac{a}{2}, \frac{b}{2}\right)$,
and infer these variance parameters from the data.

\subsection{Model Fitting Algorithm}   
The original GS-BART algorithm developed in \citet{he2026gs} does not apply to GSBCF because it only considers a single latent function included as a separate term in a (generalized) regression setting. In GSBCF, multiple latent functions are introduced; in particular, the heterogeneous treatment effect function $\tau(\cdot)$ can be viewed as a varying coefficient associated with the centered treatment variable. 
Specifically,  the GSBCF outcome regression model can be written as
$\bfy = \sum_{t=1}^{T^{(\mu)}} \boldsymbol{1} \cdot \bfg_{t}^{(\mu)} + \sum_{t=1}^{T^{(\tau)}} (\bfz - 0.5) \cdot \bfg_{t}^{(\tau)}+\bepsilon$, 
where $\bfy=\{y_i\}_{i=1}^{n}$, $\bfz=\{z_i\}_{i=1}^{n}$, $\bepsilon=\{\epsilon_i\}_{i=1}^{n}$, $\bfg_{t}^{(\mu)}=\{g_{t}^{(\mu)}(\bfx_i,\hat{e}_i, \bfs_i)\}_{i=1}^{n}$, $\bfg_{t}^{(\tau)}=\{g_{t}^{(\tau)}(\bfx_i,\bfs_i)\}_{i=1}^{n}$, and  the operator ``$\cdot$'' denotes element-wise multiplication. 
We follow a backfitting procedure to iteratively update each decision tree, $g_{t}(\cdot)$, in our latent functions conditional on the other decision tree components and $\sigma^2$. This step can be unified into a varying-coefficient model framework.  Let $\bfr_{-t}$ denote the partial residuals of $\bfy$ obtained by subtracting the current contributions of all trees except the $t$th tree. We sample $(\calT_t,\calM_t)$ and the associated leaf weight prior variance using the informed importance tempering algorithm (see Section~\ref{subsec:IIT}) under the partial-residual varying coefficient likelihood model
\begin{equation}\label{eq:partialVC}
\bfr_{-t}=\bfc_t\cdot \bfg_t(\calT_t,\calM_t)+\bepsilon
\end{equation}
In particular, the Gibbs sampling algorithm for posterior inference proceeds as follows:

 \begin{itemize}
        \item \textbf{Stage 1: Update prognostic forest.}  
        Grow $T^{(\mu)}$ trees comprising the forest for the prognostic term 
        $\mu\!\left(\bfx, \hat{e}, \bfs \right)$.  
        For each tree $\mathcal{T}^{(\mu)}_t$, 
       $\bfr_{-t}^{(\mu)}=\bfy-\sum_{j=1,j\neq t}^{T^{(\mu)}} \bfg_{j}^{(\mu)} - \sum_{t=1}^{T^{(\tau)}} (\bfz - 0.5) \cdot \bfg_{t}^{(\tau)}$,
            $\bfc_{t}^{(\mu)}=\boldsymbol{1}$. We 
        perform the following updates:
        \begin{enumerate}
            \item Update $(\mathcal{T}^{(\mu)}_t, \mathcal{M}^{(\mu)}_t)$ given 
            $\bfr_{-t}^{(\mu)}$,
            $\bfc_{t}^{(\mu)}$,
            $\sigma^2$, $\sigma_{\mu}^2$ from the partial residual varying coefficient model, in two steps: (a) $\cT^{(\mu)}_t \mid \bfr_{-t}, \sigma^2, \sigma_{\mu}^2$; (b) $\cM_t^{(\mu)} \mid \cT^{(\mu)}_t, \sigma^2, \sigma_{\mu}^2$. 
            \item Update 
            $\sigma_{\mu}^2 \mid \{\cM^{(\mu)}_t\}_{t = 1}^{T^{(\mu)}}$.
        \end{enumerate}
    
\item \textbf{Stage 2: Update treatment forest.}  
        Grow $T^{(\tau)}$ trees comprising the forest for the treatment term 
        $\tau\!\left(\bfx, \bfs \right)$.  
        For each tree $\mathcal{T}^{(\tau)}_t$, follow steps similar to those in Stage 1 to update  $(\mathcal{T}^{(\tau)}_t, \mathcal{M}^{(\tau)}_t,\sigma_{\tau}^2)$, using the partial residual varying coefficient model where 
       $\bfr_{-t}^{(\tau)}=\bfy-\sum_{t=1}^{T^{(\mu)}} \bfg_{t}^{(\mu)} - \sum_{j=1,j\neq t}^{T^{(\tau)}} (\bfz - 0.5) \cdot \bfg_{j}^{(\tau)}$,
            $\bfc_{t}^{(\tau)}=\bfz-0.5$.        

\item \textbf{Stage 3: Update residual  variance $\sigma^2$.}         
    \end{itemize}

    \noindent
The pseudo-code for the GSBCF algorithm is summarized in Supplementary 
Alg.~\ref{alg:GSB-procedure} and Alg.~\ref{app_alg:one-weak-learner}. Posterior Bayesian
inference for the CATE $\tau(\bfx,\bfs)$ at the observed data is obtained by
summing post burn-in MCMC samples of the treatment effect trees
$\{\bfg_{j}^{(\tau)}\}$. The ATE is computed by averaging $\tau(\bfx,\bfs)$ over
the empirical covariate distribution. We can also obtain
samples from the posterior predictive distribution of $\tau(\bfx,\bfs)$ at new
values of $\bfx$ and $\bfs$ by assigning each test point to the corresponding
vertices of the candidate graphs based on its covariates and spatial
information. Given posterior samples of the treatment effect trees
$\calT^{(\tau)}$, test points are recursively assigned to leaf nodes, which subsequently yield predictive draws of
the latent function $\tau(\bfx,\bfs)$ and other predictive quantities of
interest.

\subsection{Sampling of decision tree parameters}\label{subsec:IIT}
   %
We now provide the details of the sampling algorithm under the partial-residual varying coefficient likelihood model in~\eqref{eq:partialVC}. 
We define $\vartheta_{-t} = (\bfr_{-t},\bfc_{t}, \sigma_{f}^2,\sigma^2)$, where $\sigma_{f}^2$ for $f\in \{\mu,\tau\}$ denotes the leaf weight parameter prior variance of $\mu(\cdot)$ or $\tau(\cdot)$. 
Conditional on the current tree, $\calT_t$, we follow Bayesian inference to propose a new tree $\calT^{*}_t$ by performing either a split or merge move of the current tree, $\calT_t$. In the split move, one splittable leaf node of $\calT$ is randomly chosen and split into two offspring nodes by choosing a valid decision arborescence and a valid decision edge to form a graph split at the selected leaf node. A pruning move does the opposite by randomly merging two leaf nodes with the same parent node. 
However, randomly proposing such moves, as adopted in most existing Bayesian tree-based causal inference models, may lead to a very low acceptance rate and poor MCMC mixing~\citep{kim2023mixing}.

To sample $\mathcal{T}_t$ from $p(\mathcal{T}_t \mid \vartheta_{-t})$, we instead use Informed Importance Tempering (IIT)~\citep{zhou2022rapid} to grow the tree from a root. The method reweights split/merge proposals in the neighboring states of a current tree sample and yields a rejection-free MCMC method. 
Specifically, let $q(\cdot \mid \cdot)$ be the standard random-walk split/merge kernel and  $\mathcal{N}(\mathcal{T}_t)$ be the set of neighboring states associated with $q(\cdot \mid \cdot)$. 
    Each candidate $\mathcal{T}_t^* \in \mathcal{N}(\mathcal{T}_t)$ receives the informed proposal weight,
    \begin{equation}\label{eq:eta}
    \eta(\mathcal{T}_t^* \mid \mathcal{T}_t, \vartheta_{-t}) 
    = q(\mathcal{T}_t^* \mid \mathcal{T}_t) 
      h\!\left(\frac{p(\mathcal{T}_t^* \mid \vartheta_{-t})\, q(\mathcal{T}_t \mid \mathcal{T}_t^*)}
      {p(\mathcal{T}_t \mid \vartheta_{-t})\, q(\mathcal{T}_t^* \mid \mathcal{T}_t)}\right),
     \end{equation}
    where $h$ satisfies $h(x)=x\,h(1/x)$, e.g., $h(x)=\sqrt{x}$, which ensures the local balance condition
and yields a valid MCMC transition kernel~\citep{zanella2020informed}. We note that while \citet{krantsevich2023stochastic} also uses a data-informed tree proposal for standard BART-based causal inference, their method only approximates the target posterior distribution and does not ensure a valid MCMC transition kernel.

We draw a new proposal sample, $\calT_{t}^* \in \mathcal{N}(\mathcal{T}_t)$, according to the informed proposal weight function $\eta(\cdot)$. These informed proposal samples are reversible with respect to 
    $\tilde{p}(\mathcal{T}_t \mid \vartheta_{-t}) \propto p(\mathcal{T}_t \mid \vartheta_{-t})\, Z(\mathcal{T}_t, \vartheta_{-t})$,  where $Z(\calT_t, \vartheta_{-t})  = \sum_{\calT^{*}_t \in \mathcal{N}(\calT_t)} \eta\left(\calT^{*}_t | \calT_t, \vartheta_{-t}\right)$ is the normalizing constant. We can then use them as importance samples or draw a sample of $\calT_t$ using importance reweighted sampling, where the importance weight is $w(\calT^*)=1/Z(\calT^{*}_t,\vartheta_{-t})$ for each informed proposal sample $\calT^*$.  
These samples are always accepted, which avoids the low-acceptance issues of the standard tree sampling algorithm. 
    
Given a sample of $\calT_{t}$, the conditional posterior distributions of $\mu_{t,\xi}$ and $\sigma^2$ are given by 
    \begin{equation*}
      \begin{aligned}
         \mu_{t, \xi } & |  \cT_t, \bfr_{-t}, \sigma^2_f, \sigma^2 \sim \;  N \left(\frac{\bfc_{t, \xi}^\top \bfr_{-t, \xi}}{\|\bfc_{t, \xi}\|^2_2 + \frac{\sigma^2}{\sigma^2_{f}}},\; \frac{\sigma^2}{\|\bfc_{t, \xi}\|^2_2 + \frac{\sigma^2}{\sigma^2_{f}}} \right) \\
         \sigma^2 & | \{\cT_{t}, \cM_{t}\}_{t=1}^T \sim\; \text{inv-Gamma} \left(\frac{n + \kappa}{2}, \frac{ \|\bfr\|^2_2 + \eta}{2} \right)
    \end{aligned}
    \end{equation*}
    where the vector $\mathbf{r}_{-t,\xi}$ is the subvector of $\mathbf{r}_{-t}$  corresponding to observations in leaf $\xi$, that is,
    $\{r_{-t,i}: i \in \xi \}$. Similarly, $\mathbf{c}_{t,\xi}$ denotes the subvector of coefficients  $\{c_{t,i}: i \in \xi\}$.
    $\|\cdot\|_2$ denotes the Euclidean norm. $\kappa, \eta$ are hyperparameters of the inverse-Gamma prior for $\sigma^2$. Likewise, the leaf weight variance also has inverse-Gamma conjugacy, to be specific:
    \begin{equation*}
    \begin{aligned}
    \sigma_{f}^2 &\sim\;  \text{inv-Gamma} \left( \frac{ \sum_{t=1}^T | \cM_t | +   \kappa_f}{2}, \; \frac{ \sum_{t = 1 }^T \|\cM_t\|_2^2 + \eta_f}{2} \right) 
    \end{aligned}
    \end{equation*}
    where $\kappa_f$ and $\eta_f$ are hyperparameters of the inverse-gamma prior for $\sigma_{f}^2$.
    
 
The locally informed proposal weight function in~\eqref{eq:eta} involves calculating the posterior probability for every possible tree in the neighboring state:
    \begin{equation*}
    p(\mathcal{T}_t \mid \vartheta_{-t}) \;\propto\; \pi(\mathcal{T}_t)\,\prod_{\xi \in \underline{\xi}(\mathcal{T}_t)} m_{\xi}.
    \end{equation*}
where $m_{\xi}$ is the marginal partial residual data likelihood for leaf node $\xi$ of $\cT_t$ after integrating out the node-level leaf weight parameter $\mu_{t,\xi}$. 
We derive a close-form expression for $m_{\xi}$ as follows:  
\begin{equation}\label{eq:marginal-likelihood}
    \begin{aligned}
    m_{\xi}  &= \int \text{N}(\mu_{t,\xi}; 0, \sigma^2_f) 
\prod_{i \in \xi} \text{N}(r_{-t,i}; \mu_{t,\xi}, \sigma^2 / c^2_i) \, d\mu_{t,\xi} \\
& = (2 \pi \sigma^2)^{- \frac{|\xi|}{2}} 
        \times \sqrt{\frac{\sigma^2}{  \|\bfc_{t, \xi}\|^2_2 \sigma^2_f  + \sigma^2}} 
        \times \exp\left\{-\frac{1}{2 \sigma^2} \left( \|\bfr_{-t, \xi}\|^2_2 - \frac{ \left( \bfc_{t, \xi}^\top \bfr_{-t, \xi} \right)^2}{ \|\bfc_{t, \xi}\|^2_2 + \frac{\sigma^2}{\sigma^2_{f}}}\right) \right\}.
    \end{aligned}
\end{equation} 
Here $\text{N}(\cdot;\mu,\sigma^2)$ denotes the normal density with mean $\mu$ and variance $\sigma^2$.
    
If a node $\xi$ is split into children $(\xi^*_L, \xi^*_R)$, the log marginal likelihood ratio between $\mathcal{T}^*$ and $\mathcal{T}$ reduces to 
$\log m_{\xi^*_L} + \log m_{\xi^*_R} - \log m_{\xi}$. The number of possible split moves in the neighborhood of the current tree grows rapidly with the graph and tree sizes, making naive separate evaluation of marginal likelihood ratios impractical. Exploiting the ordered arborescence structure, we develop a parallel, vectorized, and recursive algorithm to efficiently compute these ratios for all candidate split moves; details are deferred to Supplementary Section \ref{appsub:rec-par-alg}.

    \section{Numerical Studies}\label{sec:simu}
    We evaluate the performance of GSBCF on various synthetic datasets and a real dataset. Candidate graph set construction methods are provided in Supplementary Section \ref{appsub:graphset}.
    
\subsection{Simulation studies}\label{subsec:sim}


\noindent\textbf{Synthetic data generation.}
We consider four true data-generating processes (DGPs). DGP1 and DGP2 are defined on a regular square lattice of areal units, illustrating GSBCF's application to areal unit data, while DGP3 and DGP4 use irregularly spaced point-referenced data on a U-shaped domain, illustrating the advantages of GSBCF for complex domains.
DGP2 and DGP4 (heterogeneous TEs) are our primary scenarios, while DGP1 and DGP3 (homogeneous TEs) are included mainly to assess the robustness of the methods under homogeneous effects. 

In DGP1 and DGP2, we simulate data on a $30\times30$ square lattice.
Eight spatial covariates are generated independently from proper CAR models using a four-nearest-neighbor adjacency matrix, with spatial autocorrelation parameters randomly sampled from $(0.9, 1.0)$ and conditional variance $\sigma=1$. Covariates $x_2$ and $x_4$ are generated as piecewise-constant functions of the spatial coordinates $\mathbf{s}_i = (s_{i1}, s_{i2})$ (see Supplementary Section \ref{appsub:x2x4} for details). 
Covariates $x_2$ and $x_4$ are unobserved spatial confounders used in data generation but excluded from model fitting. All remaining covariates, including irrelevant variables, are included in model fitting.  
Treatment assignment follows
$Z_i\sim\mathrm{Ber}(e_i)$ with
$\mathrm{logit}(e_i)=-2\cos(\pi x_{i2}x_{i3})+2(x_{i3}-0.5)^2+x_{i4}$.
Outcomes are generated as
$Y_i=\mu(\mathbf{x}_i)+\tau_i Z_i+\varepsilon_i$, $\varepsilon_i\sim\mathcal{N}(0,\sigma^2)$,
where $\mu(\mathbf{x}_i)=5\sin(\pi x_{i1}x_{i2})+10(x_{i3}-0.5)^2$.
In DGP1, $\tau_i=2$ is a constant (ATE) and $\sigma=1$; in DGP2, $\tau_i=2+2\sin(\pi x_{i1}x_{i2})+x_{i5}$ (CATE) and $\sigma=0.1$.

In DGP3 and DGP4, we sample $n=800$ spatial locations from a two-dimensional U-shaped domain $\mathcal{S}$ and generate five covariates from a Gaussian process over $\mathcal{S}$. 
The treatment assignment function follows
$\mu_{z}(\bfs,\bfx)=d_2 x_2+\tfrac{1}{5}d_1+x_1$, where $(d_1, d_2)$ are the intrinsic coordinates corresponding to spatial location $\mathbf{s}$.  Outcomes are generated as
$Y=\mu(\bfs,\bfx)+\tau(\bfs,\bfx)Z+\varepsilon$, $\varepsilon\sim\mathcal{N}(0,\sigma^2)$,
with $\mu(\bfs,\bfx)=g(\bfs)x_5+\tfrac{1}{10}x_3$, where $g(\bfs)$ is the \textsc{FELSPLINE} test function, a smooth function defined on the Ushape domain~\citep{ramsay2002spline}, and $\sigma=0.15$. In DGP3, $\tau(\bfs,\bfx)=1$ (ATE); in DGP4,
$\tau(\bfs,\bfx)=1+d_1 x_4+d_2 x_1$ (CATE).

\noindent\textbf{Competing methods.}   
    We compare the proposed GSBCF model with several benchmark Bayesian causal inference approaches:  
    (1) \textbf{BCF}, BART-based causal inference model implemented via the \textit{bcf} package in \textsf{R}~\citep{hahn2020bayesian};  
    (2) \textbf{XBCF}, implemented in the \textit{stochtree} package~\citep{krantsevich2023stochastic};  
    (3) \textbf{BART}, the standard response surface BART treating $Z$ as another covariate; 
    (4) \textbf{BART-$f_0 f_1$}, fitting separate BART models to the treatment and control data; 
    (5) \textbf{ps-BART}, propensity score is included as a predictor in BCF~\citep{hahn2020bayesian}. 
  For DGP1 and DGP3, we additionally report ATE estimates obtained using the \textit{PSweight} package~\citep{zhou2020psweight}, based on the spatial propensity scores estimated from our model. Specifically, we apply inverse probability weighting (\textbf{IPW}) and overlap weighting (\textbf{OPW}) to estimate the ATE.  There is very limited benchmark spatial causal inference software. Both \citet{osama2019inferring} and  \citet{reich2021review} do not allow for measured spatial confounders, and the latter only assumes a homogeneous treatment effect. To facilitate comparison, we adapt the spatial nearest neighbor Gaussian process regression~\citep{datta2016hierarchical} to construct a spatial Gaussian process causal inference model that incorporates the estimated propensity score via B-spline basis functions (\textbf{GP-Causal}).

\noindent\textbf{Experiment settings and evaluation metrics.}  
We set the number of trees for both the prognostic and treatment effect functions to 50 across all tree-based methods. 
GSBCF and XBCF are run for 225 MCMC iterations with a 25-iteration burn-in due to their more efficient grow-from-root and data-informed proposal schemes, which typically yield larger effective sample sizes per iteration~\citep{krantsevich2023stochastic,he2026gs}, whereas BCF and BART-based methods rely on standard split/merge samplers that generally require longer chains and are therefore run for 6000 iterations with a 2000-iteration burn-in.

Performance is evaluated using the root mean squared error (RMSE), assessing point estimation accuracy, and the continuous ranked probability score~\citep[CRPS,][]{gneiting2007strictly}, a proper scoring rule that evaluates the accuracy of the entire estimated posterior distribution. 

\begin{table}[h]
  \centering
  \caption{Comparison of ATE estimation performance under homogeneous treatment effects for DGP1 (square lattice) and DGP3 (U-shaped domain). CRPS for IPW and OPW are not available. RMSE and CRPS (and their SDs) are multiplied by 10 for DGP1 and by 100 for DGP3.}
  \label{tab:ATE-homo-combined}
  { 
  \begin{tabular}{lcccc}
    \toprule
    & \multicolumn{2}{c}{\textbf{DGP1 (Square lattice)}} 
    & \multicolumn{2}{c}{\textbf{DGP3 (U-shaped domain)}} \\
    \cmidrule(lr){2-3} \cmidrule(lr){4-5}
    \textbf{Method}
    & RMSE (SD) & CRPS (SD)
    & RMSE (SD) & CRPS (SD) \\
    \midrule
    GSBCF    & \textbf{0.88} (0.64) & \textbf{0.73} (0.28) & \textbf{1.77} (1.08) & \textbf{1.39} (0.44) \\
    BCF      & 1.85 (1.74) & 1.34 (1.27) & 2.05 (1.45) & 1.43 (0.93) \\
    XBCF     & 1.23 (1.15) & 1.00 (0.59) & 2.48 (1.47) & 1.69 (0.78) \\
    ps-BART  & 1.37 (1.10) & 1.00 (0.61) & 2.25 (2.00) & 1.67 (1.43) \\
    BART     & 1.73 (2.03) & 1.35 (1.59) & 2.21 (1.85) & 1.59 (1.38) \\
    BART-$f_0 f_1$ & 21.4 (1.58) & 1.96 (1.58) & 4.95 (3.63) & 3.63 (2.65) \\
    
    GP-Causal & 27.7 (0.83) &	2.45 (0.82)	& 1.82 (1.10) &	1.48 (0.38) \\
    IPW & 32.3 (0.99) & - & 39.8 (1.68) & - \\
    OPW & 11.5 (0.79) & - & 19.4 (1.28)   & - \\
    \bottomrule
  \end{tabular}
  }
\end{table}

\begin{table}[h]
  \centering
  \caption{Comparison of CATE estimation performance under heterogeneous treatment effects for DGP2 (square lattice) and DGP4 (U-shaped domain). RMSE and CRPS (and their SDs) are multiplied by 10 for both DGP2 and DGP4.}
  \label{tab:CATE-hetero-combined}
  { 
  \setlength{\tabcolsep}{4pt}
  \renewcommand{\arraystretch}{0.95}
  \begin{tabular}{lcccc}
    \toprule
    & \multicolumn{2}{c}{\textbf{DGP2 (Square lattice)}} 
    & \multicolumn{2}{c}{\textbf{DGP4 (U-shaped domain)}} \\
    \cmidrule(lr){2-3} \cmidrule(lr){4-5}
    \textbf{Method}
    & RMSE (SD) & CRPS (SD)
    & RMSE (SD) & CRPS (SD) \\
    \midrule
    GSBCF    & \textbf{5.00} (0.65) & \textbf{2.70} (0.28) & \textbf{5.18} (0.30) & \textbf{2.65} (0.11) \\
    BCF      & 6.80 (1.25) & 3.85 (0.64) & 7.08 (0.83) & 3.92 (0.43) \\
    XBCF     & 8.58 (0.41) & 5.20 (0.21) & 6.82 (0.50) & 3.47 (0.24) \\
    ps-BART  & 5.49 (3.93) & 2.96 (1.32) & 7.56 (0.83) & 4.38 (0.49) \\
    BART     & 7.75 (0.99) & 4.54 (0.57) & 7.18 (0.70) & 4.15 (0.39) \\
    BART-$f_0 f_1$ & 33.2 (0.32) & 14.3 (0.22) & 9.77 (0.78) & 4.97 (0.41) \\
    \bottomrule
  \end{tabular}
  }
\end{table}
\vspace{2.5em}

\noindent\textbf{Results.}
Tables~\ref{tab:ATE-homo-combined} and~\ref{tab:CATE-hetero-combined} present the estimation accuracy of treatment effects for GSBCF and all other competing methods on the synthetic datasets.
For each DGP, we repeat the experiments 50 times and report the two evaluation metrics averaged across replicates.  
For Bayesian tree-based causal inference models in DGP1 and DGP3, ATE estimates are obtained by taking the sample average of the posterior draws of the CATE. 
Across all four data-generating processes, GSBCF consistently achieves the best performance in terms of both RMSE and CRPS. Under homogeneous treatment effects (DGP1 and DGP3), GSBCF yields substantially lower ATE errors than competing methods on both the regular square lattice and the nonconvex U-shaped domain.  The improved performance in terms of CRPS indicates that GSBCF also achieves more accurate and better-calibrated uncertainty quantification. We also note that Bayesian tree-based causal models significantly outperform IPW and OPW estimators, potentially due to their flexible nonparametric outcome models.  GP-Causal performs comparably to tree-based methods in DGP3 (simple smooth surface, homogeneous effects), but performs poorly in DGP1, with substantially larger RMSE. Moreover, GP-Causal is restricted to homogeneous treatment effects and cannot be applied to DGP2 and DGP4.

\begin{figure}
    \centerline{
    \includegraphics[scale=.8]{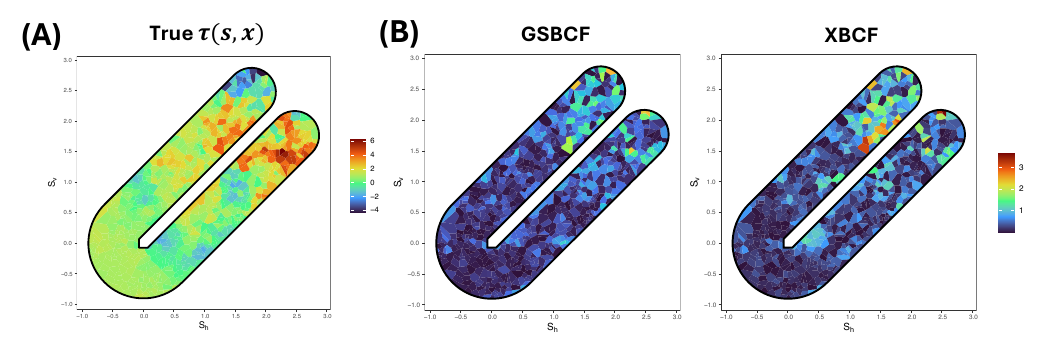}}
    \caption{(A) True spatial surface of $\tau(\textbf{s}, \textbf{x})$ in DGP4.
    (B) Mean absolute error of the estimated $\tau(\mathbf{s}, \mathbf{x})$ for GSBCF and XBCF in DGP4. 
    }
    \label{fig:Ushape results}
\end{figure}

Under heterogeneous treatment effects (DGP2 and DGP4), GSBCF again provides the most accurate CATE estimates, while other competing methods have noticeably larger errors, especially on the complex U-shaped domain. This result is not surprising, as GSBCF explicitly incorporates graph-based spatial structure and is therefore better suited to capture spatial dependence and treatment effect heterogeneity on irregular domains. As illustrated in Figure \ref{fig:Ushape results}, although both methods exhibit some degree of oversmoothing, GSBCF preserves substantially more local variation and structural detail than XBCF, resulting in a closer approximation to the true CATE surface on the U-shaped domain.

GSBCF also achieves empirical coverage of 0.96 and 0.97 for
nominal 95\% pointwise CATE credible intervals in DGP2 and DGP4,
respectively, indicating coverage close to the nominal level.
Full comparisons are provided in Supplementary
Section~\ref{appsub:cate_coverage}.


\section{Real Data Analysis of County Level Chronic Obstructive Pulmonary Disease}\label{subsec:real}
Chronic Obstructive Pulmonary Disease (COPD) is a complex, chronic respiratory condition 
    and a leading cause of morbidity and mortality in the United States, with well-documented disparities in prevalence and outcomes between rural and non-rural populations~\citep{gaffney2022health}. Understanding the causal effect of the rural–urban divide on COPD prevalence is therefore of considerable public health interest. In this study, we focus on estimating the causal effect of rurality on age-adjusted COPD prevalence across contiguous U.S. counties, excluding counties in Florida and Connecticut due to data missingness and inconsistency.
    
    The response variable $Y$ represents the 2023 age-adjusted prevalence of COPD. 
    The treatment variable is the 2023 Rural–Urban Continuum Codes (RUCC) provided by the U.S. Department of Agriculture~\citep{usda_rucc23}, with RUCC codes 1–3 classified as non-rural ($Z=1$) and 4–9 as rural ($Z=0$). Figure~\ref{app_fig:COPD-rural} displays the spatial distribution of rural and urban counties.
    Additional covariates are obtained from the CDC/ATSDR Social Vulnerability Index (SVI), 
 including 27 county-level demographic, socioeconomic, and environmental vulnerability factors. 
    Both the county-level COPD prevalence and SVI data are obtained from the CDC PLACES dataset~\citep{greenlund2022places}.
    SVI data can be broadly categorized into four groups: (1) Socioeconomic Status, (2) Household Characteristics, (3) Racial and Ethnic Minority Status, and (4) Housing Type and Transportation. A complete list of the 29 predictors, including longitude and latitude of each county's centroid location, is provided in Supplementary
    Table~\ref{app_tab:copd_covariates}.

\begin{figure}
      \centerline{
      \includegraphics[scale=.8]{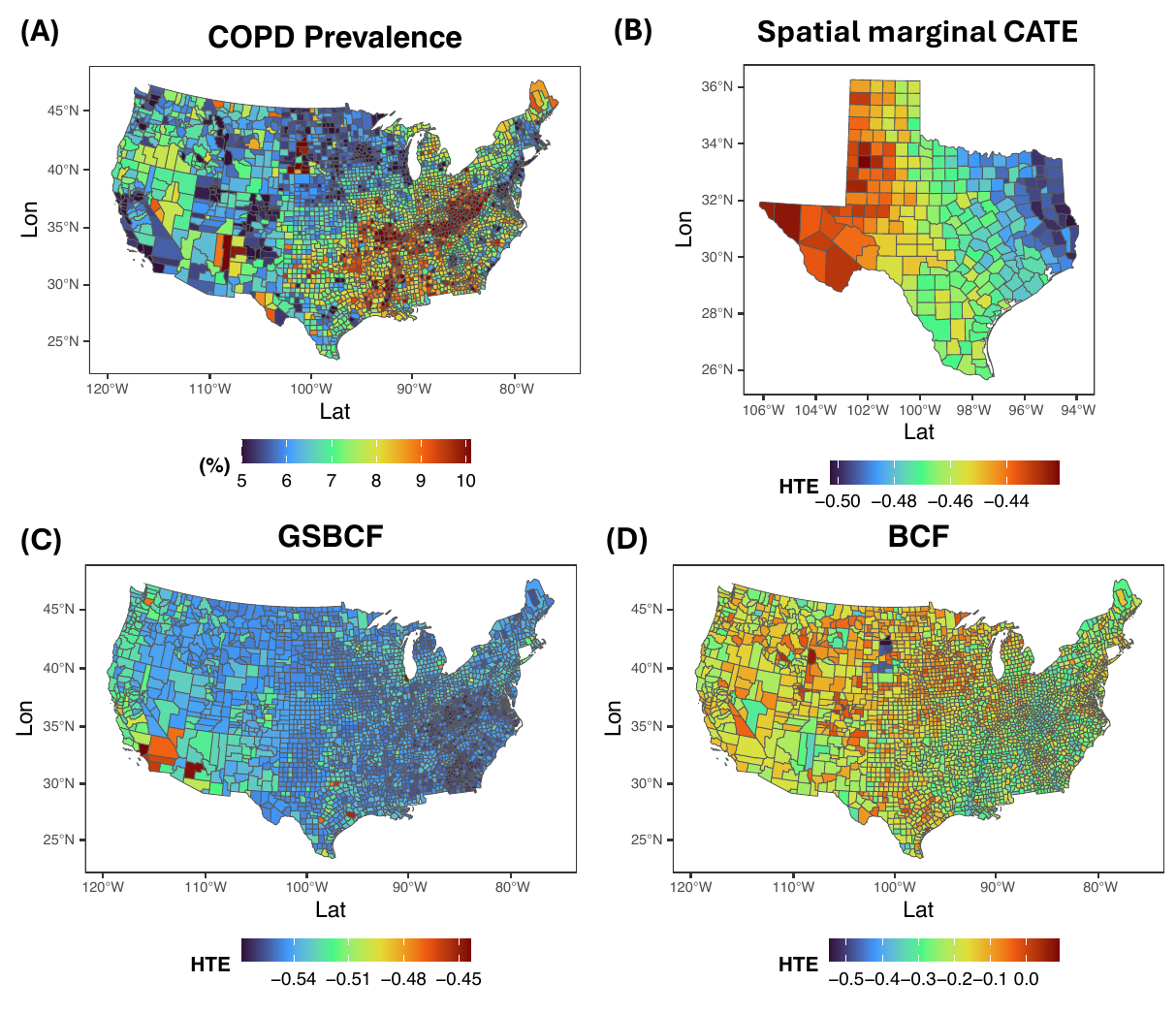}}
      \caption{(A) Spatial distribution of county-level COPD prevalence across the United States.
      (B) The spatially marginalized heterogeneous treatment effects across Texas counties for GSBCF. 
      (C) Predictive estimated heterogeneous treatment effects across U.S. counties for GSBCF.
      (D) Predictive estimated heterogeneous treatment effects across U.S. counties for BCF.}
      \label{fig:COPD-vis}
    \end{figure}

    COPD prevalence exhibits clear spatial patterns as shown in Figure~\ref{fig:COPD-vis}(A). For example, the southern and western coastal regions tend to be more urbanized than the central areas, while counties in the southeastern United States generally have higher COPD prevalence compared to other regions. 
    
\begin{figure}
  \centerline{
  \includegraphics[scale=.8]{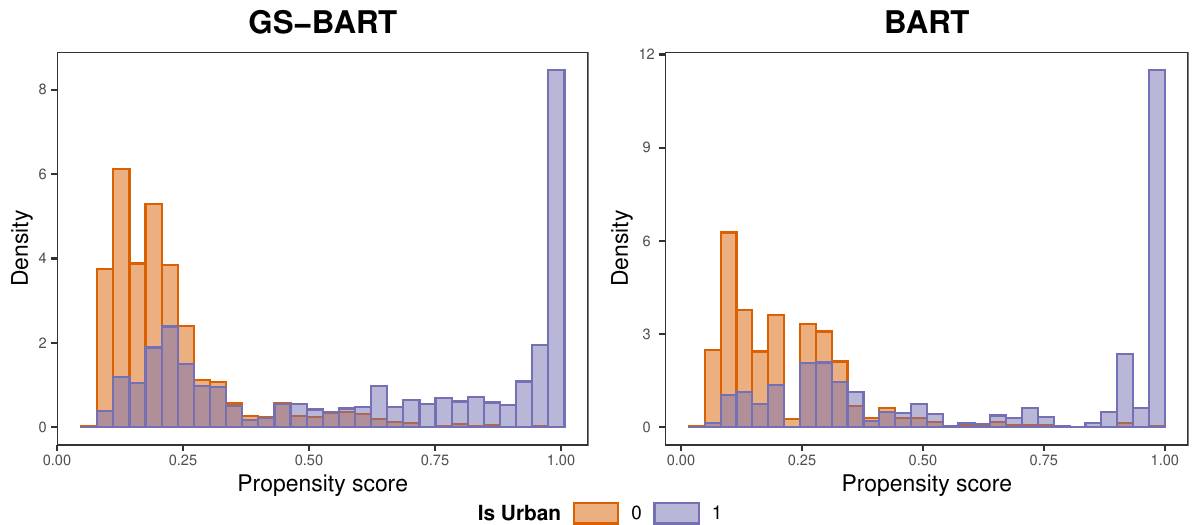}}
  \caption{Estimated propensity score distributions for urban and rural counties in the COPD dataset under GS-BART and BART.}
  \label{fig:COPD-prop}
\end{figure}

    To evaluate whether incorporating spatial effects improves CATE estimation, we compare the results of GSBCF with those of non-spatial methods such as BCF. 
    We construct a spatial neighborhood graph connecting adjacent counties and then follow the approaches described in Section~\ref{sec:GS-BCF} to construct candidate graph sets required to fit GSBCF. 
    For GSBCF and BCF, we first estimate propensity scores from using GS-BART and BART, respectively, with the binary rural indicator treatment variable as the response.
    Figure~\ref{fig:COPD-prop} shows the overlapping condition of our estimated propensity score. We observe that, for both methods, except for counties that are easily classified as urban, the remaining observations exhibit substantial overlap.

    To obtain the spatially marginalized CATE for Texas counties, we fix each county’s spatial location and evaluate the fitted model using covariate profiles from all Texas counties, forming a pseudo test dataset. The model is trained on the full dataset. After obtaining CATE predictions on the pseudo test data, we average these predictions over covariate profiles to marginalize out non-spatial effects, yielding location-specific spatially marginalized CATE.

    The ATEs estimated by GSBCF and BCF are -0.54 (95\% Credible Interval: [-0.66,-0.43]) and -0.19 (95\% Credible Interval: [-0.26,-0.13]), respectively.
    Both estimates indicate a positive causal effect of rural residence on COPD prevalence, consistent with some prior studies ~\citep[see, e.g.,][]{raju2019rural} that showed rural regions have a higher prevalence of COPD than urban regions, particularly in poorer communities that lack adequate healthcare access and health education. GSBCF yields a substantially larger ATE in magnitude, indicating a more pronounced effect. In contrast, IPW yields an estimated Urban-versus-Rural effect of (-0.84) (95\% CI: (-2.53,0.86)). 
    
    Figure~\ref{fig:COPD-vis}(C) and (D) show the estimated CATE surfaces, $\tau(\bfx_i,\bfs_i)$, from GSBCF and BCF based on the observed data.  The estimated HTEs from GSBCF exhibit a smoother and stronger spatial pattern, whereas the BCF estimates appear noisier across regions. In addition, GSBCF reveals meaningful subregional heterogeneity within Texas. The estimated rural causal effects in the Austin, Houston, and Dallas metropolitan areas are substantially smaller in magnitude than those in surrounding regions. These metropolitan areas are characterized by greater healthcare availability~\citep{ekren2025health}, which may be associated with the weaker rural--urban disparities observed in these areas. Rural counties near large metropolitan centers may also have greater access to medical infrastructure and healthcare services. 
    
    To illustrate the utility of GSBCF for investigating the functional relationship between features and treatment effect, we select Texas counties as an example to study their spatially marginalized CATE. Figure~\ref{fig:COPD-vis}(B) summarizes the spatial variation in the estimated rural--non-rural contrast across Texas. The
    spatially marginalized CATE tends to be less negative in western counties and more negative in eastern counties, which may suggest that the increase in COPD prevalence associated with rural residence
    may be more pronounced in eastern Texas. However, this pattern should be interpreted with caution because the posterior mean west-minus-east contrast was $0.05$ (95\% credible interval: $[-0.03,0.27]$), with a posterior probability of $0.66$ that the western regional average exceeds the eastern average.  
    
\section{Conclusion} 
We propose GSBCF, a graph-split decision rule–based Bayesian causal forest
framework for estimating spatially heterogeneous treatment effects in
observational studies with complex spatial structure. GSBCF flexibly captures spatial dependence and heterogeneity in treatment effects, as well as in modeling the propensity score and prognostic functions. The proposed informed
importance tempering algorithm enables highly efficient posterior exploration
over complex tree spaces.  Several extensions of GSBCF are of particular interest. One extension is to incorporate propensity-score uncertainty
more fully into posterior inference, rather than conditioning
on a fixed estimate $\hat e$. This would require jointly
addressing uncertainty propagation and computational efficiency. Although we focus on
Gaussian outcomes, the framework can be extended to non-Gaussian responses,
such as binary or count data, following the generalized regression framework.  Extending GSBCF to continuous or multi-group treatments by considering generalized propensity score regression~\citep{imai2004causal} is another interesting future work. This would require extending GS-BART from regression to nonparametric conditional density estimation. Finally, extending GSBCF to accommodate spatial
interference is an important direction for future work~\citep{giffin2023generalized}, as it would allow GSBCF to address potential violations of SUTVA in some applications. For example, one possible extension is to formulate the outcome model under a potential outcomes framework with interference. Let $Y_i(Z_i, \tilde Z_i)$ denote the potential outcome for unit $i$ under its own treatment $Z_i$ and spillover exposure $\tilde Z_i$. We consider the following propensity score regression model: $Y_i = \mu(x_i, s_i, \tilde e_i) + \tau_1(x_i, s_i)(Z_i - 0.5) + \tau_2(x_i, s_i)\tilde Z_i + \varepsilon_i$, 
where $\tau_1(x_i, s_i)$ represents the heterogeneous direct treatment effect and $\tau_2(x_i, s_i)$ captures the heterogeneous spillover (interference) effect. Both $\tau_1$ and $\tau_2$ can be assigned GS-BART priors. We leave a full development of this framework for future work.

\vspace{-2mm}
\begin{description}  
\item[Supplementary Material:] The supplementary material included below consists of  
(1) Additional details of the GSBCF algorithms; (2) Details of candidate graph set construction; (3) Additional details of synthetic datasets generation; (4) Additional simulation results; (5) Additional details and results for the COPD example.
\end{description}

\clearpage
\renewcommand{\baselinestretch}{1}\normalsize
\section*{Supplementary Materials}

\setcounter{section}{0}
\setcounter{subsection}{0}
\setcounter{equation}{0}
\setcounter{table}{0}
\setcounter{figure}{0}
\setcounter{algorithm}{0}
\setcounter{proposition}{0}
\renewcommand{\thesection}{S\arabic{section}}
\renewcommand{\thesubsection}{\thesection.\arabic{subsection}}
\renewcommand{\theequation}{S\arabic{equation}}
\renewcommand{\thetable}{S\arabic{table}}
\renewcommand{\thefigure}{S\arabic{figure}}
\renewcommand{\thealgorithm}{S\arabic{algorithm}}
\renewcommand{\theproposition}{S\arabic{proposition}}
\renewcommand{\thesubfigure}{\arabic{figure}\alph{subfigure}}
\renewcommand{\theHsection}{supp.\arabic{section}}
\renewcommand{\theHsubsection}{supp.\arabic{section}.\arabic{subsection}}
\renewcommand{\theHequation}{supp.\arabic{equation}}
\renewcommand{\theHtable}{supp.\arabic{table}}
\renewcommand{\theHfigure}{supp.\arabic{figure}}
\providecommand{\theHalgorithm}{}
\renewcommand{\theHalgorithm}{supp.\arabic{algorithm}}
\providecommand{\theHproposition}{}
\renewcommand{\theHproposition}{supp.\arabic{section}.\arabic{proposition}}

\section{Details of Model Fitting Algorithms}\label{app:algorithm}
\subsection{Hyperparameter choice}

The hyperparameters of the inverse-Gamma priors are partially data-dependent. Specifically, we follow the default prior calibration commonly used in BART \citep{chipman2010bart} and generalized GS-BART \citep{he2026gs}.
For the residual variance, we assume
$$
\sigma^2 \sim \operatorname{Inv\text{-}Gamma}
\left(\frac{\nu}{2},\frac{\nu\lambda}{2}\right).
$$
We fix the shape parameter at $\nu=3$ and select the scale parameter $\lambda$ in a data-dependent manner such that
$$
\Pr(\sigma < \hat{\sigma})=0.9,
$$
where $\hat{\sigma}$ is the sample standard deviation of the response. This follows the standard BART empirical-Bayes calibration, which places most prior mass below a rough data-based estimate of the residual scale.

For the leaf-weight variance of each latent function $f\in\{\mu,\tau\}$, we follow the generalized GS-BART prior and assume
$$
\sigma_f^2
\sim
\operatorname{Inv\text{-}Gamma}
\left(\frac{a_f}{2},\frac{b_f}{2}\right),
$$
where we fix $a_f=3$ and set
$$
b_f=\frac{\operatorname{Var}(\mathbf y)}{T_f},
$$
with $T_f$ denoting the number of trees used to model $f$. Thus, the scale hyperparameter adapts to the empirical scale of the response and the number of trees, while $\sigma_f^2$ itself is subsequently learned from the data through its posterior distribution rather than being fixed.

\subsection{Parallel and recursive algorithm}\label{appsub:rec-par-alg}
Suppose the current decision tree $\calT$ has $\ell$ leaf nodes
$\underline{\xi}=(\xi_1,\ldots,\xi_\ell)$. For a candidate arborescence
$\dG\in\dbmG$, observations associated with each vertex $v\in V(\dG)$
are grouped by leaf membership as
$(v\cap\xi_1,\ldots,v\cap\xi_\ell)$.
Consider a candidate decision edge $\de_v\in E(\dG)$ used to split a leaf
$\xi_k$. Removing $\de_v$ partitions $\dG$ into a right arborescence
$v\cup\operatorname{desc}(v)$ and a left arborescence. Denote the corresponding
offspring by $\xi^{*}_{k,R,v}$ and $\xi^{*}_{k,L,v}$, and define $\underline{\xi}^{*}_{R, v} = (\xi^{*}_{1, R, v}, \ldots, \xi^{*}_{\ell, R, v})$, and $\underline{\xi}^{*}_{L, v} = (\xi^{*}_{1, L, v}, \ldots, \xi^{*}_{\ell, L, v})$.

The log marginal likelihood ratio for each split move is  $\log m_{\xi^*_L} + \log m_{\xi^*_R} - \log m_{\xi}$, which requires computing summary statistics, $J(\xi_k)$, $H(\xi_k)$, $J(\xi^*_{k,R,v})$, $J(\xi^*_{k,L,v})$, $H(\xi^*_{k,R,v})$, and $H(\xi^*_{k,L,v})$. Here $J(\xi)=\bfc_{t, \xi}^\top \bfr_{-t, \xi}$, and $H(\xi)=\|\bfc_{t, \xi}\|^2_2$. 
To efficiently compute these summary statistics for each possible split, 
we first precompute the summary statistics of $J$ and $H$ at each vertex and leaf node. In particular, 
we define and compute the vertex-level vectors
$\underline{J}(v)=(J^1(v),\ldots,J^\ell(v))$ and
$\underline{H}(v)=(H^1(v),\ldots,H^\ell(v))$, where $
J^{k}(v) = \sum_{i\in v\cap\xi_k}
\Bigl( c_{t,i} r_{-t, i} \Bigr),\;
H^{k}(v) = \sum_{i\in v\cap\xi_k}
c^2_{t, i}, \; k=1,\ldots,\ell $.  
Aggregating over vertices yields $
\underline{J}(\underline{\xi})=\sum_{v\in V(\dG)}\underline{J}(v),\;
\underline{H}(\underline{\xi})=\sum_{v\in V(\dG)}\underline{H}(v)$. 

For a bottom edge $\de_v$, the right offspring contains only $v$, so
$\underline{J}(\underline{\xi}^{*}_{R,v})=\underline{J}(v)$ and
$\underline{H}(\underline{\xi}^{*}_{R,v})=\underline{H}(v)$. Left-offspring
statistics are computed by  $J(\xi_{L, v}^{*})=\underline{J}(\underline{\xi})-\underline{J}(v)$, and $H(\xi_{L, v}^{*})=\underline{H}(\underline{\xi})-\underline{H}(v)$. For a non-bottom edge, we recursively propagate summary statistics information
from leaves to the root as follows
$\underline{J}(\underline{\xi}^{*}_{R,v})
  = \underline{J}(v)+\sum_{u\in\operatorname{desc}(v)}\underline{J}(u),
\underline{H}(\underline{\xi}^{*}_{R,v})
  = \underline{H}(v)+\sum_{u\in\operatorname{desc}(v)}\underline{H}(u)$. 
Left-offspring quantities are obtained again by subtraction from $\underline{J}(\underline{\xi})$ and $\underline{H}(\underline{\xi})$. A split is valid if both offspring contain at least one observation and the edge is non-redundant.

We can repeat the above recursive algorithm for each candidate graph
in $\dbmG$ using parallel computation. The resulting split ratios are then combined with merge ratios, prior
ratios, and proposal probabilities to form the IIT weight function $\eta(\cdot)$. The expressions for the tree prior ratio and proposal probabilities $q(\cdot)$ are the same as those provided in~\citet{he2026gs}.

\begin{algorithm}[H]
\centering
\caption{GSBCF Gibbs sampler}\label{alg:GSB-procedure}
\small
\begin{algorithmic}
\Require Training data $(\bfy,\bfx,\bfz,\bfs,\hat{e})$, candidate graph sets $\{\dbmG^{(\mu)}_t\}_{t=1}^{T^{(\mu)}}$, $\{\dbmG^{(\tau)}_t\}_{t=1}^{T^{(\tau)}}$, number of iterations $N$.
\Ensure Posterior samples of $\mu(\cdot)$ and $\tau(\cdot)$. 

\State Initialize forests $\{\cT^{(\mu)}_t,\cM^{(\mu)}_t\}_{t=1}^{T^{(\mu)}}$, $\{\cT^{(\tau)}_t,\cM^{(\tau)}_t\}_{t=1}^{T^{(\tau)}}$ and variances $\sigma^2,\sigma_\mu^2,\sigma_\tau^2$.
\For{$j=1,\ldots,N$}

\Statex \textbf{Stage 1: Update prognostic forest $\mu(\bfX,\hat e,\bfS)$.}
\For{$t=1,\ldots,T^{(\mu)}$}
\State Compute partial residual and coefficient:
\[
\bfr_{-t}^{(\mu)}=\bfy-\sum_{k\neq t}^{T^{(\mu)}} \bfg_{k}^{(\mu)}-\sum_{k=1}^{T^{(\tau)}}(\bfz-0.5)\cdot\bfg_{k}^{(\tau)},\qquad
\bfc_{t}^{(\mu)}=\mathbf{1}.
\]
\State Update tree structure $\cT_t^{(\mu)} \mid \bfr_{-t}^{(\mu)},\bfc_{t}^{(\mu)},\sigma^2,\sigma_\mu^2$ using IIT (Alg.~\ref{app_alg:one-weak-learner}) over $\dbmG^{(\mu)}_t$.
\State Update leaf parameters $\cM_t^{(\mu)} \mid \cT_t^{(\mu)},\bfr_{-t}^{(\mu)},\bfc_{t}^{(\mu)},\sigma^2,\sigma_\mu^2$.
\State Update $\bfg_t^{(\mu)} \leftarrow g_t^{(\mu)}(\cT_t^{(\mu)},\cM_t^{(\mu)})$.
\EndFor
\State Update $\sigma_\mu^2 \mid \{\cM_t^{(\mu)}\}_{t=1}^{T^{(\mu)}}$.
\State Compute $\mu_i^{(j)} \leftarrow \sum_{t=1}^{T^{(\mu)}} g_t^{(\mu)}(\bfx_i,\hat{e}_i,\bfs_i)$ for $i=1,\ldots,n$.

\Statex \textbf{Stage 2: Update treatment forest $\tau(\bfX,\bfS)$.}
\For{$t=1,\ldots,T^{(\tau)}$}
\State Compute partial residual and coefficient:
\[
\bfr_{-t}^{(\tau)}=\bfy-\sum_{k=1}^{T^{(\mu)}} \bfg_{k}^{(\mu)}-\sum_{k\neq t}^{T^{(\tau)}}(\bfz-0.5)\cdot\bfg_{k}^{(\tau)},\qquad
\bfc_{t}^{(\tau)}=\bfz-0.5.
\]
\State Update tree structure $\cT_t^{(\tau)} \mid \bfr_{-t}^{(\tau)},\bfc_{t}^{(\tau)},\sigma^2,\sigma_\tau^2$ using IIT in Alg.~\ref{app_alg:one-weak-learner} over $\dbmG^{(\tau)}_t$.
\State Update leaf parameters $\cM_t^{(\tau)} \mid \cT_t^{(\tau)},\bfr_{-t}^{(\tau)},\bfc_{t}^{(\mu)},\sigma^2,\sigma_\tau^2$.
\State Update $\bfg_t^{(\tau)} \leftarrow g_t^{(\tau)}(\cT_t^{(\tau)},\cM_t^{(\tau)})$.
\EndFor
\State Update $\sigma_\tau^2 \mid \{\cM_t^{(\tau)}\}_{t=1}^{T^{(\tau)}}$.
\State Compute $\tau_i^{(j)} \leftarrow \sum_{t=1}^{T^{(\tau)}} g_t^{(\tau)}(\bfx_i,\bfs_i)$ for $i=1,\ldots,n$.
\Statex \textbf{Stage 3: Update residual variance.}
\State Update $\sigma^2 \mid \bfy,\{\bfg_t^{(\mu)}\}_{t=1}^{T^{(\mu)}},\{\bfg_t^{(\tau)}\}_{t=1}^{T^{(\tau)}}$.

\EndFor
\State 
Store $\{\mu_i^{(j)}\}_{j=1}^N$ and $\{\tau_i^{(j)}\}_{j=1}^N$, for $i=1,\ldots,n$.
\end{algorithmic}
\end{algorithm}

\begin{algorithm}[H] 
\centering
\caption{Informed Importance Tempering (IIT) to sample one graph-split weak learner based on the partial residual varying-coefficient (VC) model.}\label{app_alg:one-weak-learner}
\small
\begin{algorithmic}
\Require Partial-residual VC inputs $\vartheta_{-t}=(\bfr_{-t},\bfc_{t},\sigma_f^2,\sigma^2)$ with $f\in\{\mu,\tau\}$;  $\cT_t$ with only a root node; candidate graph set $\dbmG_t$; proposal kernel $q(\cdot\mid\cdot)$ with neighborhood $\mathcal{N}(\cT_t)$ induced by split/merge moves; number of informed proposals $K$.
\Ensure One draw $(\cT_t,\cM_t)\sim p(\cT_t,\cM_t\mid \vartheta_{-t})$.

\For{$k=1,\ldots,K$}
  \State Enumerate the proposal neighborhood $\mathcal{N}(\cT_t)$ under split/merge moves:
  \Statex \hspace{1em}\emph{Split:} choose a splittable leaf; choose a valid decision arborescence and decision edge from $\dbmG_t$ to form a graph split.
  \Statex \hspace{1em}\emph{Merge:} merge two leaf nodes sharing the same parent (inverse of split).
  \State For each $\cT_t^*\in\mathcal{N}(\cT_t)$, compute the informed weight
  \[
  \eta(\cT_t^*\mid \cT_t,\vartheta_{-t})
  = q(\cT_t^*\mid \cT_t)\,
    h\!\left(
    \frac{p(\cT_t^*\mid \vartheta_{-t})\,q(\cT_t\mid \cT_t^*)}
         {p(\cT_t\mid \vartheta_{-t})\,q(\cT_t^*\mid \cT_t)}
    \right).
  \]
  \State Compute the normalizing constant $Z(\cT_t,\vartheta_{-t})=\sum_{\cT_t^*\in\mathcal{N}(\cT_t)} \eta(\cT_t^*\mid \cT_t,\vartheta_{-t})$.
  \State Propose $\cT_t^{*}\sim \eta(\cdot\mid \cT_t,\vartheta_{-t})/Z(\cT_t,\vartheta_{-t})$ and set $\cT_t^{(k)}\leftarrow \cT_t^{*}$.
  \State Sample leaf parameters $\cM^{(k)}_t=\{\mu^{(k)}_{t,\xi}\}_{\xi\in\mathcal{L}(\cT_t)}$ from the conjugate Gaussian full conditional:
  \[
  \mu^{(k)}_{t,\xi}\mid \cT_t,\bfr_{-t},\sigma_f^2,\sigma^2
  \sim
  N\!\left(
  \frac{\bfc_{t,\xi}^\top \bfr_{-t,\xi}}{\|\bfc_{t,\xi}\|_2^2+\sigma^2/\sigma_f^2},\;
  \frac{\sigma^2}{\|\bfc_{t,\xi}\|_2^2+\sigma^2/\sigma_f^2}
  \right).
  \]
  \State Compute the importance weight $w^{(k)}=1/Z(\cT^{(k)}_t,\vartheta_{-t})$ and store $(\cT_t^{(k)},\cM_t^{(k)},w^{(k)})$.
\EndFor
\State Draw $(\cT_t,\cM_t)$ from $\{(\cT_t^{(k)},\cM_t^{(k)})\}_{k=1}^K$ with probability proportional to $\{w^{(k)}\}_{k=1}^K$.
\end{algorithmic}
\end{algorithm}
 
\subsection{Propensity-score estimation}
\label{appsub:propensity}

We model treatment assignment using a Bernoulli likelihood
with a logistic link:
\[
Z_i \mid e_i \sim \operatorname{Bernoulli}(e_i),
\qquad
\operatorname{logit}(e_i)
=\mu_z(\mathbf{x}_i,\mathbf{s}_i).
\]
The treatment-assignment function is represented by a GS-BART
ensemble,
\[
\mu_z(\mathbf{x}_i,\mathbf{s}_i)
=
\sum_{t=1}^{T^{(e)}}
g_t^{(e)}(\mathbf{x}_i,\mathbf{s}_i;
\mathcal{T}_t^{(e)},\mathcal{M}_t^{(e)}).
\]
The candidate graph sets incorporate both covariate chain
graphs and spatial adjacency-based arborescences, allowing
treatment assignment to depend flexibly on observed covariates,
spatial location, and their interactions.

Posterior estimation follows the generalized GS-BART algorithm
of \citet{he2026gs}. We obtain the propensity-score estimate
by averaging the treatment probabilities over the posterior:
\[
\hat e_i
=
\mathbb{E}\!\left[
\operatorname{logit}^{-1}
\{\mu_z(\mathbf{x}_i,\mathbf{s}_i)\}
\mid \mathbf{Z},\mathbf{X},\mathbf{S}
\right].
\]
Thus, averaging is performed on the probability scale,
rather than applying the inverse-logit transformation to the
posterior mean linear predictor.

The estimated scores are subsequently included as an additional
predictor in the prognostic forest through a corresponding
covariate chain graph, but are not included in the
treatment-effect forest. The primary analysis treats
$\hat e_i$ as fixed when fitting the outcome model, with
the propensity-score model estimated separately from the
outcome model.

\section{Details of candidate graph set construction}\label{appsub:graphset}
    Specifically, for unstructured numerical variables, we first sort the $n$ observations and construct an arborescence $\dG$ by binning adjacent observations. Formally, we select a positive integer $|V| \leq n$ and define cut points $-\infty = c_0 < c_1 < \dots < c_{|V|-1} < c_{|V|} = \infty$. 
    The resulting directed graph $\dG$ is a chain graph with $|V|$ vertices $\{v_1, \dots, v_{|V|}\}$, where 
    $v_k = \{i \in [n] : c_{k-1} < x_i < c_k\}$ 
    and the edge set is $\dE = \{(v_k, v_{k+1}) : 1 \leq k \leq |V| - 1\}$. 
    For the areal unit data as in DGP1, DGP2 and the real data, we construct a spatial adjacency network $A_0$, connecting two areal units if they share a boundary. We then apply a community detection algorithm~\citep{clauset2004finding} to partition the nodes in $A_0$ into $K$ connected subgraphs based on their network connectivity patterns. This procedure can be viewed as a spatial binning step on $A_0$. 
    Let $G_0$ denote the resulting binned graph, where each vertex represents a bin (i.e., a connected subgraph), and an edge is drawn between two vertices if at least one edge in $A_0$ connects nodes belonging to the corresponding subgraphs. This binned graph $G_0$ thus defines the spatial grouping structure and determines the vertex assignment of each training and testing sample.

Table~\ref{app_tab:experiment setting} reports the numbers of arborescences and their vertices used for each spatial or network-structured and unstructured feature for each numerical example.

\begin{table}[h]
    \small
    \begin{center}
    \begin{threeparttable}
    \begin{tabular}{lcc}
    \toprule[1pt]
    DATASETS & $\mathbf{|\dbmG_\text{structured}|}/\mathbf{|\dbmG_\text{unstructured}|}$ & $\mathbf{|V_\text{structured}|}/\mathbf{|V_\text{unstructured}|}$ \\
    \hline  
        DGP1 & 5/10 & 100/100  \\
        DGP2 & 5/10 & 100/100    \\ 
        DGP3  & 5/7  & 100/100  \\
        DGP4 & 5/7 & 100/100  \\
        COPD & 5/29 & 100/100   \\
    \bottomrule[1pt]
    \end{tabular} 
   \caption{The numbers of arborescences  and vertices used for the spatial or network-structured  
   and unstructured features for each experiment.}\label{app_tab:experiment setting}
    \end{threeparttable}
    \end{center}
\end{table}

\section{Details of data generation process}\label{appsub:x2x4}
The missing covariates, $x_2$ and $x_4$, have the following forms:
\begin{equation*}
    \begin{aligned}
    x_{i2} &= 0.8 
    + 1.6\,\mathbb{I}\!\left(s_{i2} > 0.7 s_{i1} + 3\right) \\
    & - 1.4\,\mathbb{I}\!\left(s_{i2} \le 0.7 s_{i1} + 3,\; s_{i2} < 18 - 0.03( s_{i1} -15)^2\right), \\
x_{i4} &= 0.2
+ 2.0\,\mathbb{I}\!\left(s_{i2} > -0.6 s_{i1} + 25\right) + \\
& 1.2\,\mathbb{I}\!\left(s_{i2} \le -0.6 s_{i1} + 25,\; s_{i2} > 12 + 3\sin\!\big((s_{i1}-2)/4\big)\right).
    \end{aligned}
\end{equation*}



\section{Additional simulation results}
\label{app:additional_simulation}

\subsection{MCMC convergence analysis} 
We assess convergence and
mixing through trace plots of posterior summaries and effective sample
sizes (ESS) based on the posterior samples of CATE RMSE.  
Each GSBCF and XBCF sweep updates both forests using informed grow-from-root proposals, typically allowing more efficient exploration of the posterior space than the local tree modifications used in conventional BCF tree updates. 

\begin{figure}[ht]
\centering
\includegraphics[width=0.88\textwidth]{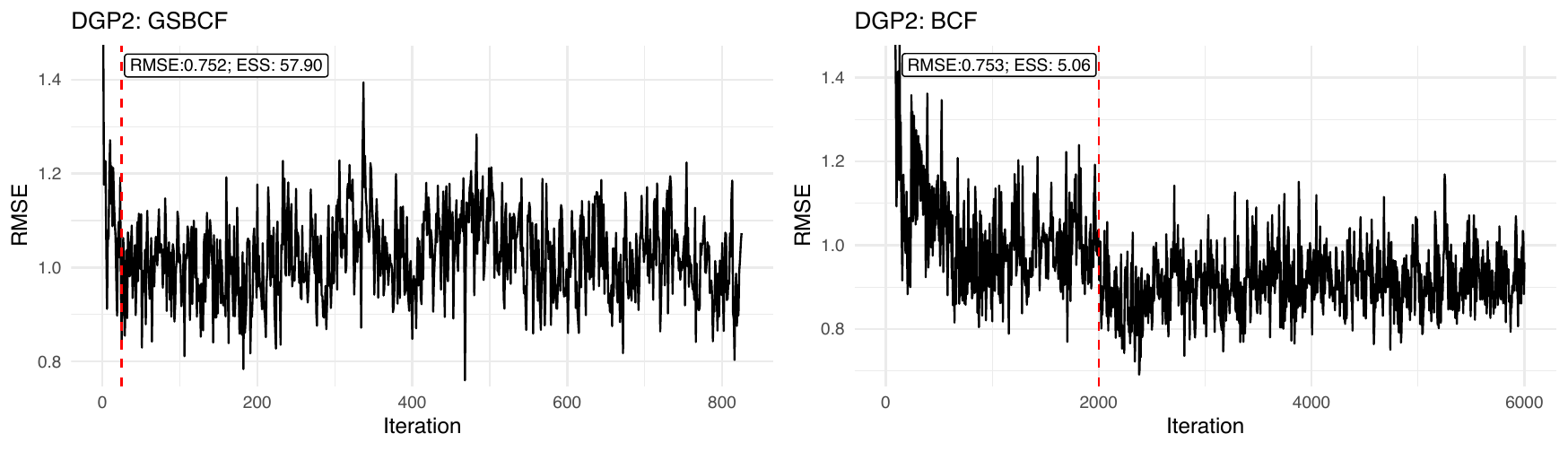}
\caption{DGP2: Post-burn-in CATE RMSE traces for one randomly selected
replication. 
Labels report posterior-mean CATE RMSE and ESS/100 for the
plotted sequence.}
\label{fig:dgp2-trace}
\end{figure}

\begin{figure}[ht]
\centering
\includegraphics[width=0.88\textwidth]{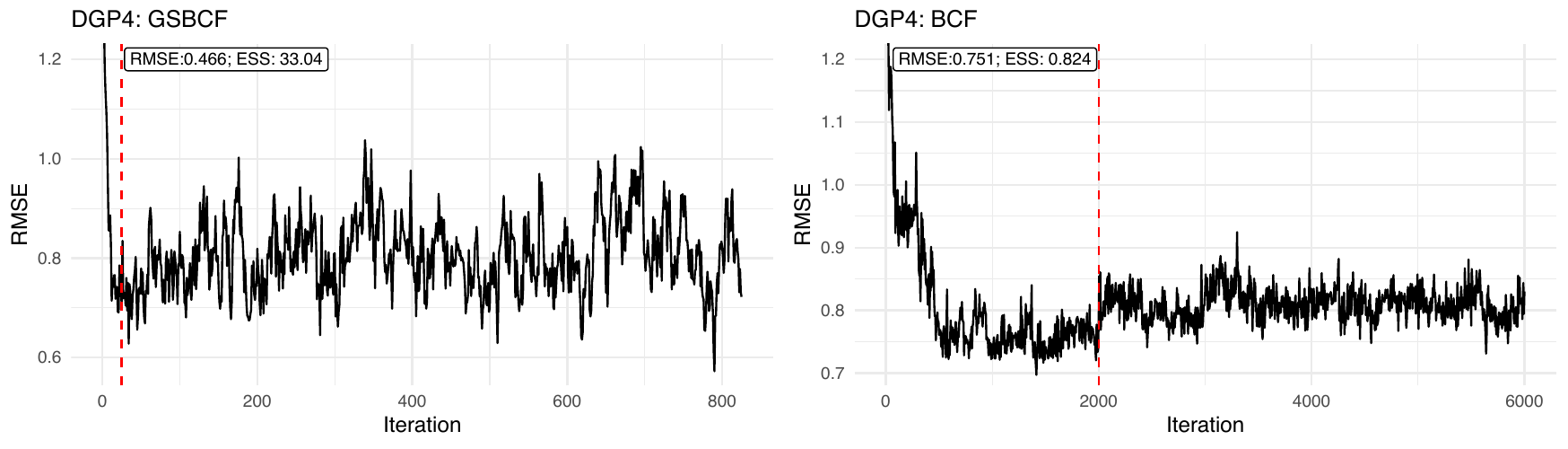}
\caption{DGP4: Corresponding diagnostics, with the same sampling
settings as Figure~\ref{fig:dgp2-trace}.}
\label{fig:dgp4-trace}
\end{figure}

Figures~\ref{fig:dgp2-trace} and~\ref{fig:dgp4-trace} show the RMSE traces for one randomly selected
replication from each of DGP2 and DGP4. 
GSBCF uses 25 burn-in and 800 post-burn-in sweeps.
BCF uses 2,000 burn-in and 4,000 post-burn-in iterations.


We also report ESS per 100 post-burn-in iterations, averaged across
simulation replications:

\begin{table}[h]
\centering
\caption{Effective sample size per 100 post-burn-in iterations
(ESS/100), averaged across simulation replications.}
\label{tab:ess_comparison}
\begin{tabular}{lcccc}
\toprule
\textbf{Method} & \textbf{DGP1} & \textbf{DGP2}
& \textbf{DGP3} & \textbf{DGP4} \\
\midrule
GSBCF & 61.23 & 33.32 & 50.96 & 40.94 \\
BCF   & 4.95  & 2.47  & 5.39  & 1.20  \\
\bottomrule
\end{tabular}
\end{table}

Overall, the RMSE traces suggest that GSBCF reaches stable post-burn-in behavior with substantially fewer burn-in iterations than BCF. Moreover,  GSBCF consistently yields higher ESS per iteration, indicating less
serial dependence and more efficient sampling.

\subsection{Sensitivity Analysis}\label{subsec:sensi}

\noindent\textit{Contribution of the propensity score.}
We conduct a Friedman-based targeted-selection experiment
\citep{hahn2020bayesian}. Let
$m_i=2\sin(\pi x_{i1}x_{i2})+5(x_{i3}-0.5)^2$ and
$t_i=|2x_{i5}-x_{i4}|$, with tildes denoting standardization
across the 900 spatial units. We generate
\[
\begin{aligned}
\mu_i &= 2\widetilde m_i,
\qquad \tau_i=0.5+0.25\tanh(\widetilde t_i),\\
e_i &= 0.05+0.90\Phi
\left(b_0+\mu_i/2+0.25\widetilde x_{i1}\right),\\
Z_i &\sim \operatorname{Bernoulli}(e_i),
\qquad Y_i=\mu_i+\tau_i Z_i+\varepsilon_i,
\end{aligned}
\]
where $\varepsilon_i\sim N(0,1)$, $\Phi$ is the standard normal
CDF, and $b_0$ sets the average treatment probability to $0.5$.
The spatial covariates $x_2$ and $x_4$ are omitted from the
fitted models. Across 50 paired replications, we regenerate
treatment and outcome and re-estimate $\widehat e$, holding
covariates, locations, and fitting settings fixed.

\begin{table}[h]
\centering
\caption{Propensity-score sensitivity analysis: mean RMSE and CRPS
with standard deviations across 50 replications.}
\label{tab:targeted_selection}
\begin{tabular}{lcc}
\toprule
\textbf{Method} & RMSE (SD) & CRPS (SD) \\
\midrule
GSBCF without $\widehat e$
& 0.271 (0.080) & 0.170 (0.060) \\
GSBCF with estimated $\widehat e$
& 0.227 (0.066) & 0.136 (0.045) \\
GSBCF with oracle $e$
& 0.183 (0.036) & 0.108 (0.022) \\
\bottomrule
\end{tabular}
\end{table}

Including $\widehat e$ reduces mean CATE RMSE and CRPS by
approximately 16\% and 20\%, respectively, with further
improvements using the true propensity score
(Table~\ref{tab:targeted_selection}). These results illustrate
the benefit of including propensity scores when treatment
assignment depends on the prognostic function.

\noindent\textit{Sensitivity to spatial interference.}
We consider an additional experiment to check the robustness of our method when SUTVA is mildly violated.
Specifically, we modify DGP2 by allowing the outcome to depend not only on a unit's own treatment, but also on a spatially weighted average of neighboring treatments. Treatment assignment follows $Z_i \sim \mathrm{Bernoulli}(e_i)$, where $
\mathrm{logit}(e_i)
=
-2\cos(\pi x_{i2}x_{i3})
+2(x_{i3}-0.5)^2
+x_{i4}.$
 
To induce mild interference, we follow \citet{giffin2023generalized} and define the normalized spillover exposure $\widetilde Z_i
=
\sum_{j \neq i} w_{ij} Z_j$,
where the weights are constructed using a Gaussian kernel:$
w_{ij}
=
\frac{\exp\{-(d_{ij}/\tau)^2\}}
{\sum_{k \neq i}\exp\{-(d_{ik}/\tau)^2\}}$, where
$d_{ij}=\|s_i-s_j\|$. We define $w_{ii}=0$. 
Thus, $\widetilde Z_i$ represents a row-normalized spatially weighted average of neighboring treatments.

The true conditional mean outcome is given by
\[
\mathbb{E}(Y_i \mid X_i, Z_i, \widetilde Z_i)
=
5\sin(\pi x_{i1}x_{i2})
+10(x_{i3}-0.5)^2
+2Z_i
+2 \widetilde Z_i,
\]  This construction introduces a moderate spillover effect while preserving the primary structure of DGP2, allowing us to assess the robustness of the proposed method under mild violations of SUTVA.

We first fit models (referred to as oracle outcome models) using the interference-adjusted response $Y_i' = Y_i - 2 \widetilde Z_i$,
which serves as an oracle baseline for comparison. We then fit GSBCF and the competing methods using the observed response $Y_i$. We refer to them as misspecified outcome models as these models ignore the interference effects of $\widetilde Z_i$.
\begin{table}[ht]
\centering
\caption{Performance comparison between oracle outcome models and misspecified outcome models under mild SUTVA violations.}
\label{tab:sutva_combined}
\begin{tabular}{lcccc}
\hline
& \multicolumn{2}{c}{Oracle Outcome} & \multicolumn{2}{c}{Misspecified Outcome} \\
Method & RMSE (SD) & CRPS (SD) & RMSE (SD) & CRPS (SD) \\
\hline
GSBCF CATE        & 5.04 (0.449) & 2.74 (0.204) & 5.57 (0.650) & 2.98 (0.228) \\
BCF CATE          & 6.95 (1.756) & 3.84 (0.791) & 6.98 (1.246) & 3.94 (0.541) \\
XBCF CATE         & 8.57 (0.353) & 5.21 (0.194) & 8.67 (0.428) & 5.24 (0.208) \\
ps-BART CATE       & 8.24 (1.228) & 4.87 (0.667) & 8.16 (1.454) & 4.74 (0.675) \\
BART CATE         & 7.88 (1.158) & 4.67 (0.652) & 8.29 (1.431) & 4.79 (0.709) \\
BART-$f_0f_1$ CATE & 33.09 (0.354) & 14.25 (0.250) & 32.98 (0.330) & 14.24 (0.228) \\
\hline
\end{tabular}
\end{table}

 The results are shown in Table~\ref{tab:sutva_combined}. RMSE and CRPS (and their standard deviations) are multiplied by 10. 
We observe that the performance of GSBCF remains robust under model misspecification, with performance only slightly worse than the oracle benchmark and still significantly outperforming other competing causal models. This robustness may arise because spillover effects are partially captured by the prognostic component that is capable of capturing the combined spatial effects through spatial graph-guided additive trees.

\subsection{Coverage of CATE credible intervals}
\label{appsub:cate_coverage}

We evaluate empirical coverage by calculating the proportion of
true unit-level CATEs contained in their nominal 95\% pointwise
credible intervals and averaging across simulation replications.
Table~\ref{app_tab:cate_coverage} shows that GSBCF achieves coverage
close to the nominal level in both DGP2 and DGP4, whereas the
competing methods exhibit undercoverage.

\begin{table}[h]
\centering
\caption{Empirical coverage of nominal 95\% pointwise CATE
credible intervals in DGP2 and DGP4, averaged across
simulation replications.}
\label{app_tab:cate_coverage}
\begin{tabular}{lcccccc}
\toprule
\textbf{DGP} & GSBCF & BCF & XBCF & ps-BART & BART
& BART-$f_0,f_1$ \\
\midrule
\textbf{DGP2} & 0.96 & 0.73 & 0.72 & 0.75 & 0.60 & 0.75 \\
\textbf{DGP4} & 0.97 & 0.66 & 0.87 & 0.53 & 0.56 & 0.76 \\
\bottomrule
\end{tabular}
\end{table}

\subsection{Performance under the Smooth and Mixed DGPs }

To assess performance
beyond piecewise-constant spatial confounding, we consider two settings
on the same $30\times30$ lattice, retaining the original observed
covariate generation process.


In the \emph{Smooth} DGP, the latent spatial fields are generated
using the squared-exponential covariance function
\[
C(s,s')=\exp\left\{-\frac{\|s-s'\|^2}{2(0.20)^2}\right\},
\]
with coordinates rescaled to $[0,1]^2$.
We obtain $U$ and $H$ by standardizing two independent GP draws,
then construct $V=0.7U+\sqrt{1-0.7^2}H$ and standardize $V$.

The \emph{Mixed} DGP adds an independent non-spatial latent
confounder to both models:
\[
\begin{aligned}
W_i &\overset{\mathrm{iid}}{\sim}N(0,1),\\
\operatorname{logit}(e_i)
&=0.4X_{i1}-0.4X_{i3}+0.7V(S_i)+0.25W_i,\\
Y_i
&=0.6X_{i1}+0.5X_{i3}+0.7U(S_i)+0.25W_i
+\tau_i(Z_i-0.5)+\epsilon_i.
\end{aligned}
\]
Here, $W_i$ is independent of the observed covariates, spatial
fields, and outcome errors. This introduces a mild violation
of the identifying assumption discussed in Section~\ref{sec:causal-framework}.


In these additional simulations and the simulations in the original manuscript, BCF, XBCF, BART, ps-BART, and
BART-$f_0,f_1$ all receive the same observed covariates and both
spatial coordinates. Table~\ref{tab:reviewer_additional_dgp} reports CATE accuracy, coverage,
ESS, and runtime. ESS is normalized to 100 original MCMC sweeps,
accounting for thinning where applicable.

\begin{table}[H]
\centering
\caption{Results under smooth spatial and mixed spatial/non-spatial
confounding. Coverage refers to 95\% CATE credible intervals;
ESS is normalized to 100 original MCMC sweeps.}
\label{tab:reviewer_additional_dgp}
\begin{tabular}{llccccc}
\toprule
DGP & Method & RMSE & CRPS & Coverage & ESS/100 & Time (s) \\
\midrule
\multirow{6}{*}{Smooth}
& GSBCF
& \textbf{0.089} & \textbf{0.050} & 0.980 & \textbf{42.75} & 772.7 \\
& BCF
& 0.105 & 0.059 & 0.956 & 6.42 & 676.2 \\
& XBCF
& 0.988 & 0.859 & 0.037 & 2.41 & 4.1 \\
& ps-BART
& 0.129 & 0.076 & 0.739 & 2.78 & 7.9 \\
& BART
& 0.128 & 0.076 & 0.730 & 3.43 & 7.7 \\
& BART-$f_0,f_1$
& 0.301 & 0.170 & 0.911 & 1.77 & 8.0 \\
\midrule
\multirow{6}{*}{Mixed}
& GSBCF
& \textbf{0.125} & \textbf{0.071} & 0.977 & \textbf{45.83} & 776.3 \\
& BCF
& 0.132 & 0.076 & 0.930 & 7.23 & 772.9 \\
& XBCF
& 1.004 & 0.869 & 0.030 & 2.58 & 4.1 \\
& ps-BART
& 0.154 & 0.093 & 0.732 & 3.00 & 7.7 \\
& BART
& 0.152 & 0.093 & 0.723 & 3.55 & 7.6 \\
& BART-$f_0,f_1$
& 0.340 & 0.192 & 0.936 & 2.32 & 7.9 \\
\bottomrule
\end{tabular}
\end{table}

GSBCF achieves the lowest RMSE and CRPS in both settings, with
slightly conservative CATE coverage. Its performance under the
Smooth DGP indicates that its advantages are not confined to
piecewise-constant confounding surfaces. Under the Mixed DGP,
estimation becomes more difficult, and the advantage over BCF
narrows. 

GSBCF also achieves higher ESS per 100 sweeps than BCF
(42.8 versus 6.4 under Smooth and 45.8 versus 7.2 under Mixed),
with comparable runtimes: 772.7 versus 676.2 seconds and
776.3 versus 772.9 seconds, respectively. These results support
improved mixing at a computational cost comparable to that of BCF,
with further scope for implementation-level optimization
of graph processing and parallel computation.
 
\section{Additional real-data details and results}
\label{appsub:realdata}

Table~\ref{app_tab:copd_covariates} lists the 29 predictors used
in the COPD analysis, comprising 27 county-level covariates
and two spatial coordinates.

\begin{table}[h]
\centering
\caption{Predictors used in the county-level COPD analysis.}
\label{app_tab:copd_covariates}
\begin{tabular}{ll}
\toprule
Variable & Description \\
\midrule
\texttt{lon} & County longitude \\
\texttt{lat} & County latitude \\

\texttt{E\_HU} & Number of housing units  \\
\texttt{E\_HH} & Number of households  \\
\texttt{E\_POV150} & Population below 150\% of the poverty level  \\
\texttt{E\_UNEMP} & Unemployed population  \\
\texttt{E\_HBURD} & Households with housing cost burden  \\
\texttt{E\_NOHSDP} & Population without a high-school diploma  \\
\texttt{E\_UNINSUR} & Uninsured population  \\

\texttt{E\_AGE65} & Population aged 65 years or older  \\
\texttt{E\_AGE17} & Population aged 17 years or younger  \\
\texttt{E\_DISABL} & Population with a disability  \\
\texttt{E\_SNGPNT} & Single-parent households  \\
\texttt{E\_LIMENG} & Population with limited English proficiency  \\

\texttt{E\_MINRTY} & Racial and ethnic minority population  \\
\texttt{E\_MUNIT} & Housing units in multi-unit structures  \\
\texttt{E\_MOBILE} & Mobile homes  \\
\texttt{E\_CROWD} & Crowded households  \\
\texttt{E\_NOVEH} & Households without a vehicle  \\
\texttt{E\_GROUPQ} & Population living in group quarters  \\

\texttt{E\_DAYPOP} & Daytime population  \\
\texttt{E\_NOINT} & Households without Internet access  \\
\texttt{E\_AFAM} & African American population  \\
\texttt{E\_HISP} & Hispanic or Latino population  \\
\texttt{E\_ASIAN} & Asian population  \\
\texttt{E\_AIAN} & American Indian and Alaska Native population  \\
\texttt{E\_NHPI} & Native Hawaiian and Pacific Islander population  \\
\texttt{E\_TWOMORE} & Population identifying as two or more races  \\
\texttt{E\_OTHERRACE} & Population identifying as another race  \\
\bottomrule
\end{tabular}
\end{table}

\begin{figure}[h]
  \centering
  \includegraphics[scale=.8]{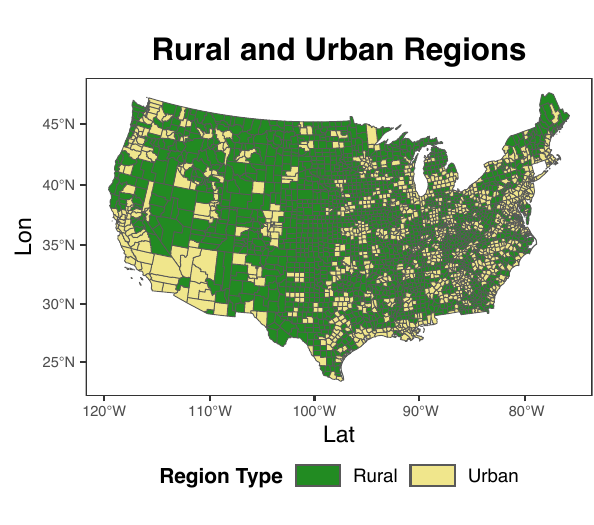}
  \caption{Spatial distribution of rural and urban counties across US.}
  \label{app_fig:COPD-rural}
\end{figure}

The spatial distribution of rural and urban counties across the US is illustrated in Figure \ref{app_fig:COPD-rural}. As shown, rural and urban counties are interspersed throughout the nation, ensuring that both region types are represented across diverse geographic locations.

\clearpage
\bibliographystyle{biom}
\bibliography{ref}

\begin{thebibliography}{}

\bibitem[\protect\citeauthoryear{Athey and Imbens}{Athey and
  Imbens}{2015}]{athey2015machine}
Athey, S. and Imbens, G.~W. (2015).
\newblock Machine learning methods for estimating heterogeneous causal effects.
\newblock {\em stat} {\bf 1050,} 1--26.

\bibitem[\protect\citeauthoryear{Chipman, George, and McCulloch}{Chipman
  et~al.}{2010}]{chipman2010bart}
Chipman, H.~A., George, E.~I., and McCulloch, R.~E. (2010).
\newblock {BART}: {B}ayesian additive regression trees.
\newblock {\em The Annals of Applied Statistics} {\bf 4,} 266--298.

\bibitem[\protect\citeauthoryear{Clauset, Newman, and Moore}{Clauset
  et~al.}{2004}]{clauset2004finding}
Clauset, A., Newman, M.~E., and Moore, C. (2004).
\newblock Finding community structure in very large networks.
\newblock {\em Physical Review E—Statistical, Nonlinear, and Soft Matter
  Physics} {\bf 70,} 066111.

\bibitem[\protect\citeauthoryear{Datta, Banerjee, Finley, and Gelfand}{Datta
  et~al.}{2016}]{datta2016hierarchical}
Datta, A., Banerjee, S., Finley, A.~O., and Gelfand, A.~E. (2016).
\newblock Hierarchical nearest-neighbor gaussian process models for large
  geostatistical datasets.
\newblock {\em Journal of the American Statistical Association} {\bf 111,}
  800--812.

\bibitem[\protect\citeauthoryear{Ekren, Maleki, Curran, Watkins, and
  Villagran}{Ekren et~al.}{2025}]{ekren2025health}
Ekren, E., Maleki, S., Curran, C., Watkins, C., and Villagran, M.~M. (2025).
\newblock Health differences between rural and non-rural texas counties based
  on 2023 county health rankings.
\newblock {\em BMC health services research} {\bf 25,} 2.

\bibitem[\protect\citeauthoryear{Fan, Li, Luo, and Sisson}{Fan
  et~al.}{2021}]{fan2021bayesian}
Fan, X., Li, B., Luo, L., and Sisson, S.~A. (2021).
\newblock Bayesian nonparametric space partitions: A survey.
\newblock In Zhou, Z.-H., editor, {\em Proceedings of the Thirtieth
  International Joint Conference on Artificial Intelligence, {IJCAI-21}}, pages
  4408--4415. International Joint Conferences on Artificial Intelligence
  Organization.
\newblock Survey Track.

\bibitem[\protect\citeauthoryear{Gaffney, Hawks, White, Woolhandler,
  Himmelstein, Christiani, and McCormick}{Gaffney
  et~al.}{2022}]{gaffney2022health}
Gaffney, A.~W., Hawks, L., White, A.~C., Woolhandler, S., Himmelstein, D.,
  Christiani, D.~C., and McCormick, D. (2022).
\newblock Health care disparities across the urban-rural divide: a national
  study of individuals with copd.
\newblock {\em The Journal of Rural Health} {\bf 38,} 207--216.

\bibitem[\protect\citeauthoryear{Gao, Wang, Stein, and Chen}{Gao
  et~al.}{2022}]{gao2022causal}
Gao, B., Wang, J., Stein, A., and Chen, Z. (2022).
\newblock Causal inference in spatial statistics.
\newblock {\em Spatial statistics} {\bf 50,} 100621.

\bibitem[\protect\citeauthoryear{Giffin, Reich, Yang, and Rappold}{Giffin
  et~al.}{2023}]{giffin2023generalized}
Giffin, A., Reich, B., Yang, S., and Rappold, A. (2023).
\newblock Generalized propensity score approach to causal inference with
  spatial interference.
\newblock {\em Biometrics} {\bf 79,} 2220--2231.

\bibitem[\protect\citeauthoryear{Gneiting and Raftery}{Gneiting and
  Raftery}{2007}]{gneiting2007strictly}
Gneiting, T. and Raftery, A.~E. (2007).
\newblock Strictly proper scoring rules, prediction, and estimation.
\newblock {\em Journal of the American statistical Association} {\bf 102,}
  359--378.

\bibitem[\protect\citeauthoryear{Greenlund, Lu, Wang, Matthews, LeClercq, Lee,
  and Carlson}{Greenlund et~al.}{2022}]{greenlund2022places}
Greenlund, K.~J., Lu, H., Wang, Y., Matthews, K.~A., LeClercq, J.~M., Lee, B.,
  and Carlson, S.~A. (2022).
\newblock Places: local data for better health.
\newblock {\em Preventing chronic disease} {\bf 19,} E31.

\bibitem[\protect\citeauthoryear{Hahn, Murray, and Carvalho}{Hahn
  et~al.}{2020}]{hahn2020bayesian}
Hahn, P.~R., Murray, J.~S., and Carvalho, C.~M. (2020).
\newblock Bayesian regression tree models for causal inference: Regularization,
  confounding, and heterogeneous effects (with discussion).
\newblock {\em Bayesian Analysis} {\bf 15,} 965--1056.

\bibitem[\protect\citeauthoryear{He, Sang, and Zhou}{He
  et~al.}{2026}]{he2026gs}
He, S., Sang, H., and Zhou, Q. (2026).
\newblock {GS-BART}: Bayesian additive regression trees with graph-split
  decision rules.
\newblock {\em Journal of the American Statistical Association} pages 1--26.

\bibitem[\protect\citeauthoryear{Imai and Van~Dyk}{Imai and
  Van~Dyk}{2004}]{imai2004causal}
Imai, K. and Van~Dyk, D.~A. (2004).
\newblock Causal inference with general treatment regimes: Generalizing the
  propensity score.
\newblock {\em Journal of the American Statistical Association} {\bf 99,}
  854--866.

\bibitem[\protect\citeauthoryear{Kane, Fang, Galetta, Goyal, Nicholson, Kepler,
  Vaccaro, and Schroeder}{Kane et~al.}{2020}]{kane2020propensity}
Kane, L.~T., Fang, T., Galetta, M.~S., Goyal, D.~K., Nicholson, K.~J., Kepler,
  C.~K., Vaccaro, A.~R., and Schroeder, G.~D. (2020).
\newblock Propensity score matching: a statistical method.
\newblock {\em Clinical spine surgery} {\bf 33,} 120--122.

\bibitem[\protect\citeauthoryear{Kim and Rockova}{Kim and
  Rockova}{2023}]{kim2023mixing}
Kim, J. and Rockova, V. (2023).
\newblock On mixing rates for bayesian cart.
\newblock {\em arXiv preprint arXiv:2306.00126} .

\bibitem[\protect\citeauthoryear{Krantsevich, He, and Hahn}{Krantsevich
  et~al.}{2023}]{krantsevich2023stochastic}
Krantsevich, N., He, J., and Hahn, P.~R. (2023).
\newblock Stochastic tree ensembles for estimating heterogeneous effects.
\newblock In {\em International Conference on Artificial Intelligence and
  Statistics}, pages 6120--6131. PMLR.

\bibitem[\protect\citeauthoryear{Li, Morgan, and Zaslavsky}{Li
  et~al.}{2018}]{li2018balancing}
Li, F., Morgan, K.~L., and Zaslavsky, A.~M. (2018).
\newblock Balancing covariates via propensity score weighting.
\newblock {\em Journal of the American Statistical Association} {\bf 113,}
  390--400.

\bibitem[\protect\citeauthoryear{Lunceford and Davidian}{Lunceford and
  Davidian}{2004}]{lunceford2004stratification}
Lunceford, J.~K. and Davidian, M. (2004).
\newblock Stratification and weighting via the propensity score in estimation
  of causal treatment effects: a comparative study.
\newblock {\em Statistics in medicine} {\bf 23,} 2937--2960.

\bibitem[\protect\citeauthoryear{Osama, Zachariah, and Sch{\"o}n}{Osama
  et~al.}{2019}]{osama2019inferring}
Osama, M., Zachariah, D., and Sch{\"o}n, T.~B. (2019).
\newblock Inferring heterogeneous causal effects in presence of spatial
  confounding.
\newblock In {\em International Conference on Machine Learning}, pages
  4942--4950. PMLR.

\bibitem[\protect\citeauthoryear{Papadogeorgou, Choirat, and
  Zigler}{Papadogeorgou et~al.}{2019}]{papadogeorgou2019adjusting}
Papadogeorgou, G., Choirat, C., and Zigler, C.~M. (2019).
\newblock Adjusting for unmeasured spatial confounding with distance adjusted
  propensity score matching.
\newblock {\em Biostatistics} {\bf 20,} 256--272.

\bibitem[\protect\citeauthoryear{Raju, Keet, Paulin, Matsui, Peng, Hansel, and
  McCormack}{Raju et~al.}{2019}]{raju2019rural}
Raju, S., Keet, C.~A., Paulin, L.~M., Matsui, E.~C., Peng, R.~D., Hansel,
  N.~N., and McCormack, M.~C. (2019).
\newblock Rural residence and poverty are independent risk factors for chronic
  obstructive pulmonary disease in the united states.
\newblock {\em American journal of respiratory and critical care medicine} {\bf
  199,} 961--969.

\bibitem[\protect\citeauthoryear{Ramsay}{Ramsay}{2002}]{ramsay2002spline}
Ramsay, T. (2002).
\newblock Spline smoothing over difficult regions.
\newblock {\em Journal of the Royal Statistical Society Series B: Statistical
  Methodology} {\bf 64,} 307--319.

\bibitem[\protect\citeauthoryear{Reich, Yang, Guan, Giffin, Miller, and
  Rappold}{Reich et~al.}{2021}]{reich2021review}
Reich, B.~J., Yang, S., Guan, Y., Giffin, A.~B., Miller, M.~J., and Rappold, A.
  (2021).
\newblock A review of spatial causal inference methods for environmental and
  epidemiological applications.
\newblock {\em International Statistical Review} {\bf 89,} 605--634.

\bibitem[\protect\citeauthoryear{Saha, Basu, and Datta}{Saha
  et~al.}{2023}]{saha2023random}
Saha, A., Basu, S., and Datta, A. (2023).
\newblock Random forests for spatially dependent data.
\newblock {\em Journal of the American Statistical Association} {\bf 118,}
  665--683.

\bibitem[\protect\citeauthoryear{Sigrist}{Sigrist}{2022}]{sigrist2022gaussian}
Sigrist, F. (2022).
\newblock Gaussian process boosting.
\newblock {\em Journal of Machine Learning Research} {\bf 23,} 1--46.

\bibitem[\protect\citeauthoryear{{USDA ERS}}{{USDA ERS}}{2023}]{usda_rucc23}
{USDA ERS} (2023).
\newblock Rural--urban continuum codes.
\newblock U.S. Department of Agriculture, Economic Research Service.

\bibitem[\protect\citeauthoryear{Zanella}{Zanella}{2020}]{zanella2020informed}
Zanella, G. (2020).
\newblock Informed proposals for local {MCMC} in discrete spaces.
\newblock {\em Journal of the American Statistical Association} {\bf 115,}
  852--865.

\bibitem[\protect\citeauthoryear{Zhan and Datta}{Zhan and
  Datta}{2024}]{zhan2024neural}
Zhan, W. and Datta, A. (2024).
\newblock Neural networks for geospatial data.
\newblock {\em Journal of the American Statistical Association} pages 1--21.

\bibitem[\protect\citeauthoryear{Zhang, Shih, and M{\"u}ller}{Zhang
  et~al.}{2007}]{zhang2007spatially}
Zhang, S., Shih, Y.-C.~T., and M{\"u}ller, P. (2007).
\newblock A spatially-adjusted bayesian additive regression tree model to merge
  two datasets.
\newblock {\em Bayesian Analysis} {\bf 2,} 611--634.

\bibitem[\protect\citeauthoryear{Zhou and Smith}{Zhou and
  Smith}{2022}]{zhou2022rapid}
Zhou, Q. and Smith, A. (2022).
\newblock Rapid convergence of informed importance tempering.
\newblock In {\em International Conference on Artificial Intelligence and
  Statistics}, pages 10939--10965. PMLR.

\bibitem[\protect\citeauthoryear{Zhou, Tong, Li, and Thomas}{Zhou
  et~al.}{2020}]{zhou2020psweight}
Zhou, T., Tong, G., Li, F., and Thomas, L.~E. (2020).
\newblock Psweight: an r package for propensity score weighting analysis.
\newblock {\em arXiv preprint arXiv:2010.08893} .

\end{thebibliography}
\label{lastpage}
\clearpage
\end{document}